\PassOptionsToPackage{arxiv}{melba}

\documentclass[twoside,11pt]{article}

\usepackage{melba}

\usepackage{mwe} 

\usepackage{framed,multirow}

\usepackage{amssymb}
\usepackage{latexsym}
\usepackage{amsmath}
\usepackage{graphicx}
\usepackage{textcomp}
\usepackage{booktabs}
\usepackage{color}
\usepackage{multirow}
\usepackage{array}
\usepackage{multicol}
\usepackage{layouts}
\usepackage{braket}
\usepackage{siunitx}
\usepackage{subfigure}
\usepackage{float}
\usepackage{bm}
\usepackage[linesnumbered,ruled,vlined]{algorithm2e}
\usepackage{url}
\usepackage{xcolor}

\usepackage{hyperref}

\melbaid{YYYY:NNN}  
\melbaauthors{Zhang, Zhao, Xue, Kellman, Klein, Tao}  
\firstpageno{1}  
\melbayear{2024}  
\datesubmitted{m1/yyyy}  
\datepublished{m2/yyyy}  

\ShortHeadings{RIIR: Recurrent Inference Image Registration}{Zhang, Zhao, Xue, Kellman, Klein and Tao}

\title{Recurrent Inference Machine for Medical Image Registration}

\author{\firstname Yi \surname Zhang \email y.zhang-43@tudelft.nl \\  
	\addr Department of Imaging Physics, Delft University of Technology, Delft, The Netherlands
	\AND
	\name Yidong Zhao \email y.zhao-8@tudelft.nl \\
	\addr Department of Imaging Physics, Delft University of Technology, Delft, The Netherlands        \AND
        \name Hui Xue \email xueh@microsoft.com \\
        \addr Health Futures, Microsoft Research,  Redmond, Washington, USA
        \AND
        \name Peter Kellman \email kellmanp@nhlbi.nih.gov\\
        \addr  National Heart, Lung, and Blood Institute, National Institutes of Health, Bethesda, Maryland, USA
        \AND
        \name Stefan Klein \email s.klein@erasmusmc.nl \\
        \addr Department of Radiology and Nuclear Medicine, Erasmus University Medical Center, Rotterdam, the Netherlands
        \AND
        \name Qian Tao \email q.tao@tudelft.nl \\
        \addr Department of Imaging Physics, Delft University of Technology, Delft, The Netherlands}

\begin{document}

\maketitle


\begin{abstract}
 Image registration is essential for medical image applications where alignment of voxels across multiple images is needed for qualitative or quantitative analysis. With recent advances in deep neural networks and parallel computing, deep learning-based medical image registration methods become competitive with their flexible modelling and fast inference capabilities. However, compared to traditional optimization-based registration methods, the speed advantage may come at the cost of registration performance at inference time. Besides, deep neural networks ideally demand large training datasets while optimization-based methods are training-free. To improve registration accuracy and data efficiency, we propose a novel image registration method, termed Recurrent Inference Image Registration (RIIR) network. RIIR is formulated as a meta-learning solver for the registration problem in an iterative manner. RIIR addresses the accuracy and data efficiency issues, by learning the update rule of optimization, with implicit regularization combined with explicit gradient input.

 
We {extensively evaluated RIIR} on brain MRI, {lung CT, and} quantitative cardiac MRI datasets, in terms of both registration accuracy and training data efficiency. Our experiments showed that RIIR outperformed a range of deep learning-based methods, even with only $5\%$ of the training data, demonstrating high data efficiency. Key findings from our ablation studies highlighted the important added value of the hidden states introduced in the recurrent inference framework for meta-learning. Our proposed RIIR offers a highly data-efficient framework for deep learning-based medical image registration.
\end{abstract}



\section{Introduction and Related Works}


Medical image registration, the process of establishing anatomical correspondences between two or more medical images, finds wide applications in medical imaging research, including imaging feature fusion \citep{haskins2020deep,oliveira2014medical}, treatment planning \citep{staring2009registration, king2010registering,byrne2022varian}, and longitudinal patient studies \citep{sotiras2013deformable,jin2021predicting}. Medical image registration is traditionally formulated as an optimization problem, which aims to solve a parameterized transformation in an iterative manner \citep{klein2007evaluation}. Typically, the optimization objective consists of two parts: a similarity term that enforces the alignments between images, and a regularization term that imposes smoothness constraints. Due to the complexity of non-convex optimization, traditional methods often struggle with long run time, especially for large, high-resolution images. This hinders its practical use in clinical practice, e.g. surgery guidance \citep{sauer2006image}, where fast image registration is demanded\citep{avants2011reproducible,balakrishnan2019voxelmorph}. 

With recent developments in machine learning, the data-driven deep-learning paradigm has gained popularity in medical image registration \citep{rueckert2019model}. Instead of iteratively updating the transformation parameters by a conventional optimization pipeline, deep learning-based methods make fast image-to-transformation predictions at inference time. Early works learned the transformation in a supervised manner \citep{miao2016cnn,yang2016fast}, while unsupervised learning methods later became prevalent. They adopt similar loss functions as those in conventional methods but optimize them through amortized neural networks \citep{balakrishnan2019voxelmorph,de2019deep}. These works demonstrate the great potential of deep learning-based modes for medical image registration. Nonetheless, one-step inference of image transformation is in principle a difficult problem, compared to the iterative approach, especially when the deformation field is large. In practice, the one-step inference requires a relatively large amount of data to train the deep learning network for consistent prediction, and may still lead to unexpected transformations at inference time \citep{fechter2020one,hering2019mlvirnet,zhao2019recursive}.

In contrast to one-step inference, recent studies revisited iterative registration, using multi-step inference processes \citep{fechter2020one,kanter2022flexible,qiu2022embedding,sandkuhler2019recurrent,zhao2019recursive}. Some of these iterative methods \cite{kanter2022flexible} \cite{qiu2022embedding} fall within the realm of meta-learning. Instead of learning the optimized parameters, meta-learning focuses on learning the optimization process itself. The use of meta-learning in optimization, as explored by \cite{andrychowicz2016learning} and \cite{finn2017model} for image classification tasks, has led to enhanced generalization and faster convergence. For medical imaging applications, a prominent example is the recurrent inference machine (RIM) by \cite{putzky2017recurrent}, originally proposed to solve inverse problems with explicit forward physics models. RIM has demonstrated excellent performance in fast MRI reconstruction \citep{lonning2019recurrent} and MR relaxometry \citep{sabidussi2021recurrent}. 

In this study, we propose a novel meta-learning medical image registration method, named Recurrent Inference Image Registration (RIIR). RIIR is inspired by RIM, but significantly extends its concept to solve more generic optimization problems: different from inverse problems, medical image registration presents a high-dimensional optimization challenge with no closed-form forward model. Below we provide a detailed review to motivate our work.



\subsection{Related Works}
In this section, we review deep learning-based medical image registration methods in more detail, categorizing them into one-step methods for direct image-to-transformation inference, and iterative methods for multi-step inference. Additionally, we provide a brief overview of meta-learning for medical imaging applications.

\subsubsection{One-step Deep Learning-based Registration}
Early attempts of utilizing convolutional neural networks (CNNs) for medical image registration supported confined transformations, such as SVF-Net \citep{rohe2017svf}, Quicksilver \citep{YANG2017378}, and the work of \cite{cao2017deformable}, which are mostly trained in a supervised manner. With the introduction of U-Net architecture \citep{ronneberger2015unet}, which has excellent spatial expression capability thanks to its multi-resolution and skip connection, \cite{balakrishnan2019voxelmorph,dalca2019unsupervised,hoopes2021hypermorph} proposed unsupervised deformable registration frameworks. In the work of \cite{de2019deep}, a combination of affine and deformable transformations was further considered. More recent methods extended the framework by different neural network backbones such as transformers \citep{zhang2021learning} or implicit neural representations \citep{wolterink2022implicit,van2023deformable}. 

\subsubsection{Iterative Deep Learning-based Registration}
However, a one-step inference strategy may struggle when predicting large and complex transformations \citep{hering2019mlvirnet,zhao2019recursive}. In contrast to one-step deep learning-based registration methods, recent work adopted iterative processes, reincarnating the conventional pipeline of optimization for medical image registration, either in terms of image resolution \citep{hering2019mlvirnet,mok2020large,fechter2020one,xu2021multi,liu2021learning}, {multiple optimization steps} \citep{zhao2019recursive,sandkuhler2019recurrent,falta2022learning,kanter2022flexible}, or combined \citep{qiu2022embedding}. In \cite{sandkuhler2019recurrent}, the use of RNN with gated recurrent unit (GRU) \citep{chung2014empirical} was considered, where each step progressively updates the transformation by adding an independent parameterized transformation. Another multi-step method proposed in \cite{zhao2019recursive} uses recursive cascaded networks to generate a sequence of transformations, which is then composed to get the final transformation. However, the method requires independent modules for each step, which can be memory-inefficient.\cite{hering2019mlvirnet} proposed a variational method on different levels of resolution, where the final transformation is the composition of the transformations from coarse- to fine-grained. \cite{fechter2020one} addresses the importance of data efficiency of deep learning-based models by evaluating the model performance when data availability is limited, and a large domain shift exists. {\mbox{\cite{falta2022learning}} proposed an iterative method named Learn-to-Optimise (L2O) to emulate the gradient-based optimization in lung CT registration. Unlike the fully unsupervised training scheme, the method utilizes a deep supervision strategy on the generated key points with a recurrent use of U-Net. Noticeably, the method uses additional input feature modalities including dynamically sampled coordinates and MIND features \mbox{\citep{heinrich2012mind}} to enhance the model. }A more recent work proposed in \cite{qiu2022embedding}, Gradient Descent Network for Image Registration (GraDIRN), integrates multi-step and multi-resolution for medical image registration. Specifically, the update rule follows the idea of conventional optimization by deriving the gradient of the similarity term \textit{w.r.t.} the current transformation and using a CNN to estimate the gradient of the regularization term. Though the direct influence of the gradient term shows to be minor compared to the CNN output \citep{qiu2022embedding}, the method bridges gradient-based optimization and deep learning-based methods. The method proposed in \cite{kanter2022flexible} used individual long short-term memory (LSTM) modules for implementing recurrent refinement of the transformation. However, the scope of the work is limited to affine transformation, which only serves as an initialization for the conventional medical image registration pipeline.

\subsubsection{Meta-Learning and Recurrent Inference Machine}
Meta-learning, also described as ``learning to learn", is a subfield of machine learning. In this approach, an outer algorithm updates an inner learning algorithm, enabling the model to adapt and optimize its learning strategy to achieve a broader objective. For example, in a meta-learning scenario, a model could be trained on a variety of tasks, such as different types of image recognition, with the goal of quickly adapting to unseen similar tasks, like recognizing new kinds of objects not included in the original training set, using a few training samples. \citep{hospedales2021meta}. An early approach in meta-learning is designing an architecture of networks that can update their parameters according to different tasks and data inputs \citep{schmidhuber1993neural}. The work of \cite{cotter1990fixed} and \cite{younger1999fixed} further show that a fixed-weight RNN demonstrates flexibility in learning multiple tasks. More recently, methods learning an optimization process with RNNs were developed and studied in \cite{andrychowicz2016learning,chen2017learning,finn2017model}, demonstrating superior convergence speed and better generalization ability for unseen tasks. 

In the spirit of meta-learning, RIM was developed by \cite{putzky2017recurrent} to solve inverse problems. RIM learns a single recurrent architecture that shares the parameters across all iterations, with internal states passing through iterations \citep{putzky2017recurrent}. 
In the context of meta-learning, RIM distinguishes two tasks of different levels: the `inner task', which focuses on solving a specific inverse problem (e.g., superresolution of an image), and the `outer task', aimed at optimizing the optimization process itself. This setting enables RIM to efficiently learn and apply optimization strategies to complex problems. {Therefore, RIM only has one neural network component which learns the outer task.} RIM has shown robust and competitive performance across different application domains, from cosmology \citep{morningstar2019data, modi2021cosmicrim} to medical imaging \citep{karkalousos2022assessment,lonning2019recurrent,putzky2019rim, sabidussi2021recurrent,sabidussi2023dtirim}. {To the best of our knowledge, most applications} of RIM aim to solve an inverse problem with a known differentiable forward model in closed form, such as Fourier transform with sensitivity map and sampling mask in MRI reconstruction \citep{lonning2019recurrent}. 
 
However, the definition of an explicit forward model does not exist for the medical image registration task. {Although RIM does not require a forward model by its design, the absence of a concrete forward model makes the problem more complicated.} {In this case, our formulation is similar to a realization of iterative amortized inference \mbox{\citep{marino2018iterative}}. In \mbox{\cite{marino2018iterative}}, a variational auto-encoder (VAE) framework is studied to learn the amortized optimization process given the input data and approximate posterior gradients where the likelihood under certain forward model could be absent.} In this work, we sought to extend the framework of RIM, which demonstrated state-of-the-art performance in medical image reconstruction challenges \citep{muckley2021results,putzky2019rim,zbontar2018fastmri}, to the medical image registration problem {which relaxes the need of gradient likelihood under specific forward model solely}. The same formulation can be generalized to other high-dimensional optimization problems where explicit forward models are absent but differentiable evaluation metrics are available.

\subsection{Contributions}

The main contributions of our work are three-fold:

1. We propose a novel meta-learning framework, RIIR, for medical image registration. RIIR learns the optimization process, in the absence of explicit forward models. RIIR is flexible \textit{w.r.t.} the input modality while demonstrating competitive accuracy in different medical image registration applications.


2. Unlike existing iterative deep learning-based methods, our method integrates the gradient information of input images into the prediction of dense incremental transformations. As such, RIIR largely simplifies the learning task compared to one-step inference, significantly enhancing the overall data efficiency, as demonstrated by our experiments.


3. Through in-depth ablation experiments, we not only showed the flexibility of our proposed method with varying input choices but also investigated how different architectural choices within the RIM framework impact its performance. In particular, we showed the added value of hidden states in solving complex optimization problems in the context of medical imaging, which was under-explored in existing literature. 
\begin{figure*}[htbp]
    \centering
    \includegraphics[scale=1.0]{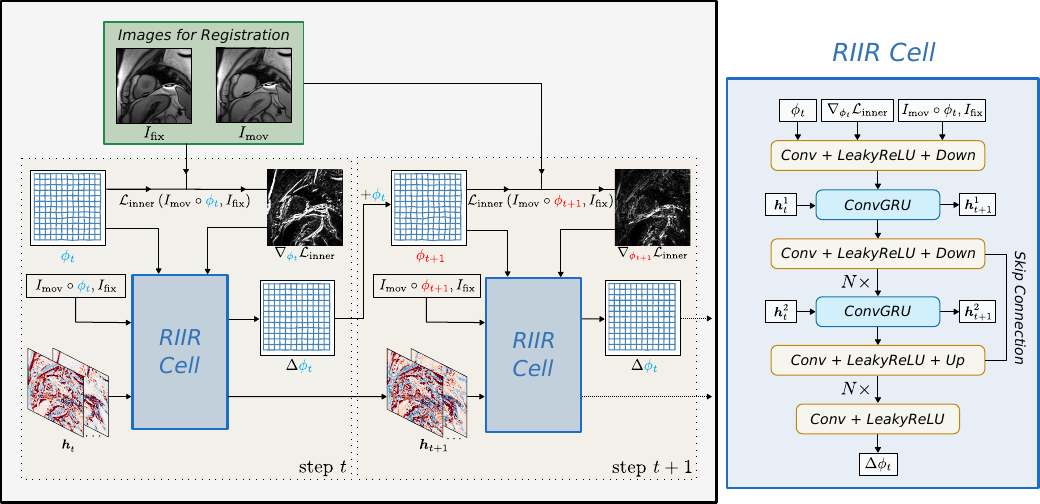}
    \caption{{Overview of RIIR framework}. Here, an illustrative cardiac image pair is shown as an example. The hidden states $\bm{h}_t = [\bm{h}_t^1,\bm{h}_t^2]$ are visualized in  channel-wise fashion. The inner loss $\mathcal{L}_{\text{inner}}$ is calculated during each step of RIIR thus dynamically changing. When $t = 0$, the deformation field $\bm{\phi}_0$ is initialized as an identical transformation. In RIIR Cell, the dimensions of Conv and ConvGRU layer are dependent on the input (2D or 3D).}
    \label{fig: RIIR example}
\end{figure*}
\section{Methods}
\subsection{Deformable Image Registration}
Deformable image registration aims to align a moving image $I_\text{mov}$ to a fixed image $I_\text{fix}$ by determining a transformation $\bm{\phi}$ acting on the shared coordinates {$\boldsymbol{\chi}$}, such that the transformed image $I_\text{mov} \circ \bm{\bm{\phi}}$ is similar enough to $I_\text{fix}$. The similarity is often evaluated by a scalar-valued metric. In deformable image registration, $\bm{\bm{\phi}}$ is considered to be a relatively small displacement added to the original coordinate {$\boldsymbol{\chi}$, expressed as $\bm{\phi} = \boldsymbol{\chi} + u(\boldsymbol{\chi})$}. Since the transformation $\bm{\phi}$ is calculated between the pair $(I_\text{mov}, I_{\text{fix}})$, the process is often referred to as \textit{pairwise} registration \citep{balakrishnan2019voxelmorph}. Finding such transformation $\bm{\phi}$ in pairwise registration can be viewed as the following optimization problem:
\begin{equation}
\hat{\bm{\phi}}=\underset{\bm{\phi}}{\operatorname{argmin}} \ \mathcal{L}_{\text {sim }}(I_\text{mov} \circ \bm{\phi}, I_\text{fix})+\lambda \mathcal{L}_{\text {reg }}(\bm{\phi}),
\label{eq: phi optimization}
\end{equation}
where $\mathcal{L}_\text{sim}$ is a similarity term between the deformed image $I_\text{mov} \circ \bm{\phi}$ and fixed image $I_{\text{fix}}$, $\mathcal{L}_\text{reg}$ is a regularization term constraining $\bm{\phi}$, and $\lambda$ is a trade-off weight term. 


\subsection{Recurrent Inference Machine (RIM)}

The idea of RIM originates from solving a closed-form inverse problem \citep{putzky2017recurrent}: \begin{equation}
    \bm{y} = A\bm{x} +\bm{n},
    \label{eq:inv_problem}
\end{equation}
where $\bm{y} \in \mathbb{R}^{m}$ is a noisy measurement vector, $\bm{x} \in \mathbb{R}^d$ is the underlying noiseless signal, $A \in \mathbb{R}^{m\times d}$ is a measurement matrix, and $\bm{n}$ is a random noise vector. When $m \ll d$, the inverse problem is ill-posed. Thus, to constrain the solution space of $x$, a common practice is to solve a \textit{maximum a posteriori} (MAP) problem:
\begin{equation}
\max _{\bm{x}} \log \mathcal{L}_{\text{likelihood}}(\bm{y} | \bm{x})+\log p_\text{prior}(\bm{x}),
\label{eq:map}
\end{equation}
where $\mathcal{L}_{\text{likelihood}}(\bm{y}|\bm{x})$ is a likelihood term representing the noisy forward model, such as the Fourier transform with masks in MRI reconstruction \cite{putzky2019rim}, and $p_\text{prior}$ is the prior distribution of the underlying signal $\bm{x}$. A simple iterative scheme at step $t$ for solving Eq. \ref{eq:map} is via gradient descent:
\begin{equation}
    \bm{x}_{t+1} = \bm{x}_{t} + \gamma_t\nabla_{\bm{x}_t}\left(\log \mathcal{L}_{\text{likelihood}}(\bm{y}|\bm{x}) + \log p_{\text{prior}}(\bm{x})\right),
    \label{eq:map_iteration}
\end{equation}
where $\gamma_t$ denotes a scalable step length and $\nabla_{\bm{x}_t}$ denotes the gradient \textit{w.r.t.} $\bm{x}$, evaluated at $\bm{x}_t$. Then, in RIM implementation, Eq. \ref{eq:map_iteration} is represented as:
\begin{equation}
    \bm{x}_{t+1} = \bm{x}_{t} + g_{\theta}\left( \nabla_{\bm{x}_t}\left(\log \mathcal{L}_{\text{likelihood}}(\bm{y}|\bm{x})\right), \bm{x}_t\right),
    \label{eq:rim}
\end{equation}
where $g_{\theta}$ is a neural network parameterized by $\theta$. In RIM, the prior distribution {on the data prior (regularization)} $p_{\text{prior}}(\bm{x})$ is implicitly integrated into the parameterized neural network $g_{\theta}$ which is trained with a weighted sum of the individual prediction losses between $\bm{x}$ and $\bm{x}_t$ (\textit{e.g.}, the mean squared loss) at each time step $t$. 

In the context of meta-learning, we regard the likelihood term $\mathcal{L}_{\text{likelihood}}$ guided by the forward model as the `inner loss', denoted by $\mathcal{L}_{\text{inner}}$ as it is serving as the input of the neural network $g_\theta$. For example, given the Gaussian assumption of the noise $\bm{n}$ with a known variance of $\sigma^2$ and linear forward model described in Eq. \ref{eq:inv_problem}, the inner loss can be given as the logarithm of the maximum likelihood estimation (MLE) solution:
\begin{equation}
    \mathcal{L}_{\text{inner}} = \frac{1}{\sigma^2}\Vert\bm{y}-A\bm{x}\Vert_2^2.
\end{equation}
In RIM, the gradient of $\mathcal{L}_{\text{inner}}$ is calculated explicitly with the (linear) forward operator $A$, which is free of the forward pass of a neural network. That means $\mathcal{L}_{\text{inner}}$ does not directly contribute to the update of the network parameters $\theta$. The weighted loss for training the neural network $g_\theta$ for efficient solving the inverse problem can be regarded as the `outer loss', denoted by $\mathcal{L}_{\text{outer}}$. In the form of the inverse problem shown in Eq. \ref{eq:inv_problem}, the outer loss to update the network parameter $\theta$ across $T$ time steps can be expressed as:

\begin{equation}
    \mathcal{L}_{\text{outer}}(\theta) = \frac{1}{T}\sum_{i=1}^{T} \Vert \bm{x} - \bm{x}_t\Vert_2^{2}.
\end{equation}

For clarity and consistency, these notations of $\mathcal{L}_{\text{inner}}$ and $\mathcal{L}_{\text{outer}}$ will be uniformly applied in the subsequent sections.

\subsection{Recurrent Inference Image Registration Network (RIIR)}

Inspired by the formulation of RIM and the optimization nature of medical image registration, we present a novel deep learning-based image registration framework, named the Recurrent Inference Image Registration Network (RIIR). The overview of our proposed framework can be found in Fig. \ref{fig: RIIR example}. 
 
 Originally, RIM aimed to learn a recurrent solver for an inverse problem where the forward model from signal to measurement is known {for inverse problems}, such as quantitative mapping \citep{sabidussi2021recurrent} or MRI reconstruction \citep{lonning2019recurrent}. {Similarly to RIM in other medical image applications, the regularization is proposed to be learned implicitly in the neural network. Therefore}, we determine the inner loss $\mathcal{L}_{\text{inner}}$ by adapting the optimization objective in Eq. \ref{eq: phi optimization}. Specifically, we use the similarity part $\mathcal{L}_{\text{sim}}$ in Eq. \ref{eq: phi optimization} as the inner loss at time step $t$: 
 \begin{equation}
     \mathcal{L}_{\text{inner}}\left(I_\text{mov}\circ\bm{\phi}_t,I_\text{fix}\right) = \mathcal{L}_{\text{sim}}\left(I_\text{mov}\circ\bm{\phi}_t,I_\text{fix}\right). 
     \label{eq:inner loss}
 \end{equation} {The gradient of $\mathcal{L}_{\text{inner}}$ can be calculated using auto differentiation.}

 {Nevertheless, in the standard RIM framework on the optimization objective in Eq. {\ref{eq: phi optimization}}, the displacement fields serve as the primary optimization signal which may overlook the rich structural and contextual information present in the original and warped images. This information loss can be critical, particularly for capturing implicit regularization introduced by the structures in original images, which directly influences $\mathcal{L}_{\text{outer}}$ and the goal of optimizing the evaluation metric (\textit{e.g.}, the overlapping of tissues or organs) which is usually not equivalent to $\mathcal{L}_{\text{outer}}$. We recognize such potential loss of information when relying solely on the gradient of the inner loss \textit{w.r.t.} to $\bm{\phi}$.  To address this, we extend the RIM framework, drawing inspiration from a more generalized formulation: the iterative amortized inference {\mbox{\citep{marino2018iterative}}}, ensuring that the model leverages the original images more effectively in the iterative optimization process. Such an extension takes the warped images as input for optimization, while not changing the optimization objective but enriching the information that could be passed into the neural network.}
 
With this modification, our proposed framework performs an end-to-end iterative prediction of a dense transformation $\bm{\phi}$ in $T$ steps for pairwise registration: Given the input image pair $\left(I_\text{mov},I_\text{fix}\right)$, the optimization problem in Eq. \ref{eq: phi optimization} can be solved by the iterative update of $\bm{\phi}$. And the update rule at step $t \in \{0,1,\ldots, T-1\}$ is:
\begin{equation}
    \bm{\phi}_{t+1} = \bm{\phi}_t + \Delta \bm{\phi}_t,
    \label{eq: phi update}
\end{equation}
where $\bm{\phi}_0$ is initialized as an identity mapping $\bm{\phi}_0(\boldsymbol{\chi}) = \boldsymbol{\chi}$. The update at step $t$, $\Delta \bm{\phi}_t$, is calculated by a recurrent update network $g_\theta$ by taking a channel-wise concatenation of
\begin{equation}
    \{\bm{\phi}_t,\nabla_{\bm{\phi}_t}\mathcal{L}_{\text{inner}}\left(I_\text{mov}\circ\bm{\phi}_t,I_\text{fix}\right),I_\text{mov}\circ\bm{\phi}_t, I_\text{fix}\}
\end{equation}
 as input, where $\nabla_{\bm{\phi}_t}$ denotes the gradient \textit{w.r.t.} $\bm{\phi}$ evaluated for $\bm{\phi} = \bm{\phi}_t$ and $\mathcal{L}_{\text{inner}}$ denotes the inner loss.
 
 In the implementation of RIM, the iterative update Eq. \ref{eq: phi update} is achieved by a recurrent neural network (RNN) to generalize the update rule in Eq. \ref{eq:rim} with hidden memory state variable $\bm{h}$ estimated for each time step $t$, {which is the only neural network component in the whole pipeline.} Unlike previous RIM-based works \citep{putzky2019rim,sabidussi2021recurrent} which use two linear gated recurrent units (GRU) to calculate the hidden states $\bm{h}_t$, in RIIR, two convolutional gated recurrent units (ConvGRU) \citep{shi2015convolutional} are used to better preserve spatial correlation in the image. We further investigate the necessity of including such two-level recurrent structures in our experiment, particularly considering potential complexities in constructing computation graphs for neural networks. The iterative update equations of RIIR at step $t$ have the following form, with the hidden memory states:
\begin{align}
    \left\{ \Delta \bm{\phi}_t,  \boldsymbol{h}_{t+1} \right\} &= g_\theta(\bm{\phi}_t, \nabla_{\bm{\phi}_t}\mathcal{L}_{\text{inner}}\left(I_\text{mov}\circ\bm{\phi}_t,I_\text{fix}\right),I_\text{mov}\circ\bm{\phi}_t, I_\text{fix}, \boldsymbol{h}_t),\\
    \bm{\phi}_{t+1} &= \bm{\phi}_t + \Delta \bm{\phi}_t,
    \label{update riir}
\end{align}
where $\boldsymbol{h}_t = \{\bm{h}_t^1, \bm{h}_t^2\}$ denotes the two-level hidden memory states at step $t$. The size of $\bm{h}_t$ depends on the size of input image pair $(I_{\text{mov}},I_{\text{fix}})$ with multiple channels. For $t = 1$, $\bm{h}_1$ is initialized to a zero input. We name our network $g_\theta$ as RIIR Cell, with its detailed architecture illustrated in Fig. \ref{fig: RIIR example}. To address the difference between our RIIR from the existing gradient-based iterative algorithm (GraDIRN) \citep{qiu2022embedding} under the same definition of $\mathcal{L}_\text{inner}$ as in Eq. \ref{eq:inner loss}, RIIR uses the gradient of inner loss as the neural network input to calculate the incremental update. On the other hand, GraDIRN takes the channel-wise warped image pair $(I_{\text{mov}}\circ \phi, I_{\text{fix}})$ and deformation field $\phi$ as the input to the network to output regularization update in Eq. \ref{eq: phi optimization}, while the gradient of $\mathcal{L}_{\text{inner}}$ is added to the update {of the deformation} without any further processing { thus does not directly affect the network part of the pipeline}.

Unlike previous work in deep learning-based iterative deformable image registration methods which does not incorporate internal hidden states \citep{zhao2019recursive,fechter2020one,qiu2022embedding}, we propose to combine the gradient information and hidden states as the network input. {Our method also differs from \mbox{\cite{falta2022learning}} in several aspects: in \mbox{\cite{falta2022learning}}, the input consists of a collection of images, displacement, sampled coordinates and MIND features; also, the part of the U-Net output channels serve as hidden states, instead of using dedicated ConvGRU units as in our design.} Using $\boldsymbol{h}_t$ also suggests an analogy with gradient-based optimization methods such as the Limited-memory Broyden–Fletcher–Goldfarb–Shanno algorithm (L-BFGS) to track and memorize progression \citep{putzky2017recurrent}. To substantiate this design, the input selections of RIIR will be further ablation-studied and discussed in our experiments.

Since the ground-truth deformation field is not known in deformable image registration, we use the optimization objective Eq. \ref{eq: phi optimization} as the proposed outer loss to optimize the parameters $\theta$ of RIIR Cell $g_\theta$. We incorporate a weighted sum of losses for the outer loss $\mathcal{L}_{\text{outer}}$ to ensure that each step contributes to the final prediction:
\begin{equation}
\mathcal{L}_{\text {outer}}(\theta)=\sum_{t=1}^T w_t \left(\mathcal{L}_{\text {sim }}(I_\text{mov} \circ \bm{\phi}_t, I_\text{fix})+\lambda \mathcal{L}_{\text {reg }}(\bm{\phi}_t)\right),
\label{outer loss}
\end{equation}
where $w_t$ is a (positive) scalar indicating the weight of step $t$. In our experiment, both uniform ($w_t = \frac{1}{T}$) and exponential weights ($w_t = 10^{\frac{t-1}{T-1}}$) are considered and will be compared in the experiments. It is noticeable that the design of using a (weighted) average of the stepwise loss also makes our proposed RIIR different from other iterative deep learning-based methods \citep{qiu2022embedding,zhao2019recursive} which use only the final output to calculate the loss, {and \mbox{\cite{falta2022learning}} use a uniform weight,} addressing the fact that early steps in the prediction process were neglected before.

\subsection{Metrics}
\noindent \textbf{Similarity Functions for Inner Loss $\mathcal{L}_{\text{inner}}$: } 
In the context of image registration, unlike inverse problems with straightforward forward models, the problem is addressed as a broader optimization challenge. Therefore, it requires an investigation of choosing a (differentiable) function acting as the inner loss function evaluating the quality of estimation of $\bm{\phi}_t$ iteratively in RIIR. Furthermore, the gradient of $\mathcal{L}_{\text{inner}}$ as an input of a convolutional recurrent neural network has not been studied before for deformable image registration. These motivate the study on the different choices of $\mathcal{L}_{\text{inner}}$ under a fixed choice of outlet loss $\mathcal{L}_{\text{outer}}$. In this work, we evaluate three similarity functions: mean squared error (MSE), normalized cross-correlation (NCC)~\citep{avants2008symmetric}, and normalized mutual information (NMI)~\citep{studholme1999overlap}.

The MSE between two 3D images $I_1, I_2 \in \mathbb{R}^{d_x \times d_y \times d_z}$ is defined as follows:
\begin{equation}
    \operatorname{MSE} \left(I_1, I_2\right) = \frac{1}{d_xd_yd_z} {\left\| I_1 - I_2 \right\|}_2^2,
    \label{eq:mse}
\end{equation}
where $|\Omega_I| = d_xd_yd_z$ denotes the all possible coordinates. The MSE metric is minimized when pixels of $I_1$ and $I_2$ have the same intensities. Therefore, it is sensitive to the contrast change. In comparison, the NCC metric measures the difference between images with the image intensity normalized. The NCC difference between $I_1$ and $I_2$ is given by:
\begin{equation}
\operatorname{NCC}(I_1, I_2) = \frac{1}{|\Omega_{I_1}|}\sum_{\boldsymbol{\chi} \in \Omega_{I_1}} \frac{\sum_{\boldsymbol{\chi}' \in \Omega_{\boldsymbol{\chi}}}(I_1(\boldsymbol{\chi}') - \bar{I_1}(\boldsymbol{\chi}))(I_2(\boldsymbol{\chi}') - \bar{I_2}(\boldsymbol{\chi}))}{\sqrt{\hat{I}_1(\boldsymbol{\chi})\hat{I}_2(\boldsymbol{\chi})}},
\label{eq:ncc}
\end{equation}
where $\Omega_{I_1}$ denotes all possible coordinates in $I_1$, $\Omega_{\boldsymbol{\chi}}$ represents a neighborhood of voxels around coordinate position $\boldsymbol{\chi}$ and $\bar{I}(\boldsymbol{\chi})$ and $\hat{I}(\boldsymbol{\chi})$ denote the (local) mean and variance in $\Omega_{\boldsymbol{\chi}}$. 

Compared to MSE and NCC, NMI is shown to be more robust when the linear relation of signal intensities between two images does not hold~\citep{studholme1999overlap,de2020mutual}, which is often the case in quantitative MRI as the signal models are mostly exponential \citep{messroghli2004modified,chow2022improved}. The NMI between two images can be written as:
\begin{equation}
\operatorname{NMI}(I_1, I_2)=\frac{H(I_1)+H(I_2)}{H(I_1,I_2)},
\label{eq:nmi}
\end{equation}
where $H(I_1)$ and $H(I_2)$ are marginal entropies of $I_1$ and $I_2$, respectively, and $H(I_1,I_2)$ denotes the joint entropy of the two images. 
Since the gradient is both necessary for $\mathcal{L}_{\text{inner}}$ and $\mathcal{L}_{\text{outer}}$ we adopt a differentiable approximation of the joint distribution proposed in \cite{qiu2021learning} based on Parzen window with Gaussian distributions~\citep{thevenaz2000optimization}. 

\noindent \textbf{Regularization Metrics: }
To ensure a smooth and reasonable deformation field, we primarily use a diffusion regularization loss which penalizes large displacements in $\bm{\phi}$ acting on $I\in \mathbb{R}^{d_x \times d_y \times d_z}$ {\mbox{\citep{fischer2002fast}}}:
\begin{equation}
    \mathcal{L}_{\text{diff}} = \frac{1}{|\Omega_I|} \sum_{\boldsymbol{\chi} \in \Omega_{I}}\left\Vert\nabla \bm{\phi}(\boldsymbol{\chi})\right\Vert_2^2,
    \label{eq:reg}
\end{equation}
where $|\Omega_I| = d_xd_yd_z$, $\nabla \bm{\phi}(\boldsymbol{\chi})$ denotes the Jacobian of $\bm{\phi}$ at coordinate $\boldsymbol{\chi}$. It is noticeable that Eq.~\ref{eq:reg} and its gradient are not evaluated in each RIIR inference step as indicated in Eq. \ref{eq:inner loss}, the outer loss $\mathcal{L}_{\text{outer}}$ and the data-driven training process can guide the RIIR Cell $g_\theta$ to learn the regularization implicitly.

{Since RIIR aims to learn implicit regularization from data with the outer loss, we also include two additional regularization metrics in our ablation study: curvature regularization and linear elastic regularization with fixed elasticity parameters \mbox{\citep{fischer2004unified}}. The curvature loss penalizes the second spatial derivatives of the displacement field $\bm{\phi}$, encouraging smoothness in the rate of change of $\bm{\phi}$:}

\begin{equation} \mathcal{L}_{\text{curv}} = \frac{1}{|\Omega_I|} \sum_{\boldsymbol{\chi} \in \Omega_{I}} \sum_{i,j=1}^3 \left\Vert \frac{\partial^2 \bm{\phi}(\boldsymbol{\chi})}{\partial \chi_i \partial \chi_j} \right\Vert_2^2, \label{eq:curv} \end{equation} 
{where $\frac{\partial^2 \bm{\phi}(\boldsymbol{\chi})}{\partial \chi_i \partial \chi_j}$ denotes the second-order partial derivative of $\bm{\phi}$ \textit{w.r.t.} dimensions $\chi_{i}$ and $\chi_{j}$ at coordinates $\bm{\chi}$.}

{The linear elastic regularization aims to regularize the displacement field by considering both the divergence and strain of the field. Its variational formulation can be described as follows \mbox{\citep{fischer2004unified}}:}
\begin{equation} \mathcal{L}_{\text{elas}} = \frac{1}{|\Omega_I|} \sum_{\boldsymbol{\chi} \in \Omega_I} \left( \frac{\lambda_e}{2} \left( \operatorname{div}  \bm{\phi}(\boldsymbol{\chi}) \right)^2 + \frac{\mu}{4} \sum_{i,j=1}^3 \left\Vert \frac{\partial \phi_i(\boldsymbol{\chi})}{\partial \chi_j} + \frac{\partial \phi_j(\boldsymbol{\chi})}{\partial \chi_i} \right\Vert_2^2 \right), \label{eq:elastic} \end{equation} 
{where $\operatorname{div} \bm{\phi}(\boldsymbol{\chi}) = \sum_{i=1}^3 \frac{\partial \phi_i(\boldsymbol{\chi})}{\partial \chi_i}$ denotes the divergence of the displacement field, $\lambda_e$ and $\mu_e$ are Lam\'{e} parameters controlling the strength of volumetric and shear deformation respectively. The first term penalizes volume changes, while the second term regularizes shear deformation of the displacement field $\bm{\phi}$.}

\section{Experiments}
\subsection{Dataset} We evaluated our proposed RIIR framework on two separate datasets:  1) A 3D brain MRI image dataset with inter-subject registration setup, OASIS \citep{marcus2007open} with pre-processing from \cite{hoopes2021hypermorph}, denoted as \textbf{OASIS}. {2) A 3D lung CT dataset with intra-subject registration setup, National Lung Screening Trial (NLST)\mbox{\citep{aberle2011reduced}}, provided and processed by the Learn2Reg challenge \mbox{\citep{hering2022learn2reg}}. This dataset is denoted by \textbf{NLST}.} 3) A 2D quantitative cardiac MRI image datasets based on multiparametric SAturation-recovery single-SHot Acquisition (mSASHA) image time series \citep{chow2022improved}, denoted as \textbf{mSASHA}. These datasets, each serving our interests in inter-subject tissue alignment and respiratory motion correction with {and without} contrast variation.

\begin{figure}[htb!]
    \centering
    \includegraphics[scale = 0.26]{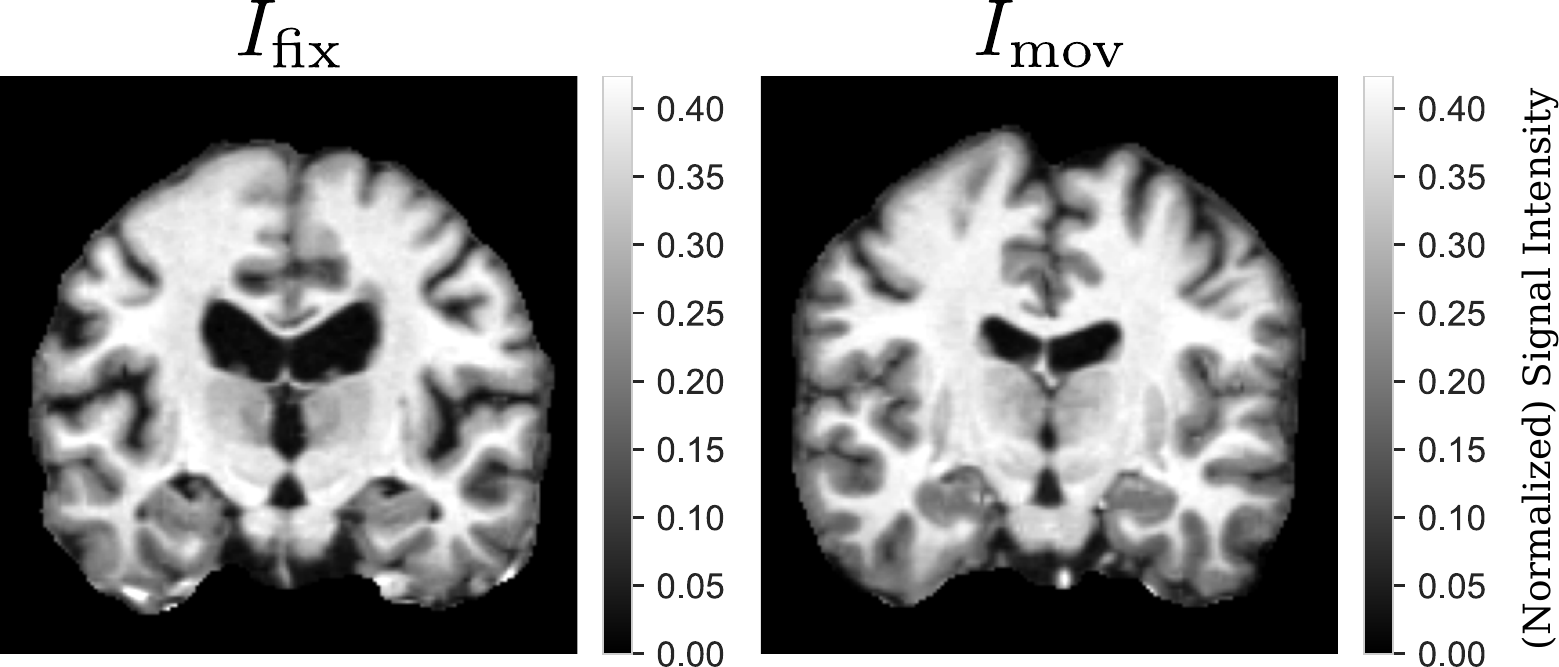}
    \caption{An example of OASIS dataset for two subjects as $I_{\text{fix}}$ and $I_{\text{mov}}$. The choices of $I_{\text{fix}}$ and $I_{\text{mov}}$ are random during training.}
    \label{fig:oasis_example}
\end{figure}

\begin{figure}[htb!]
    \centering
    \includegraphics[scale = 0.26]{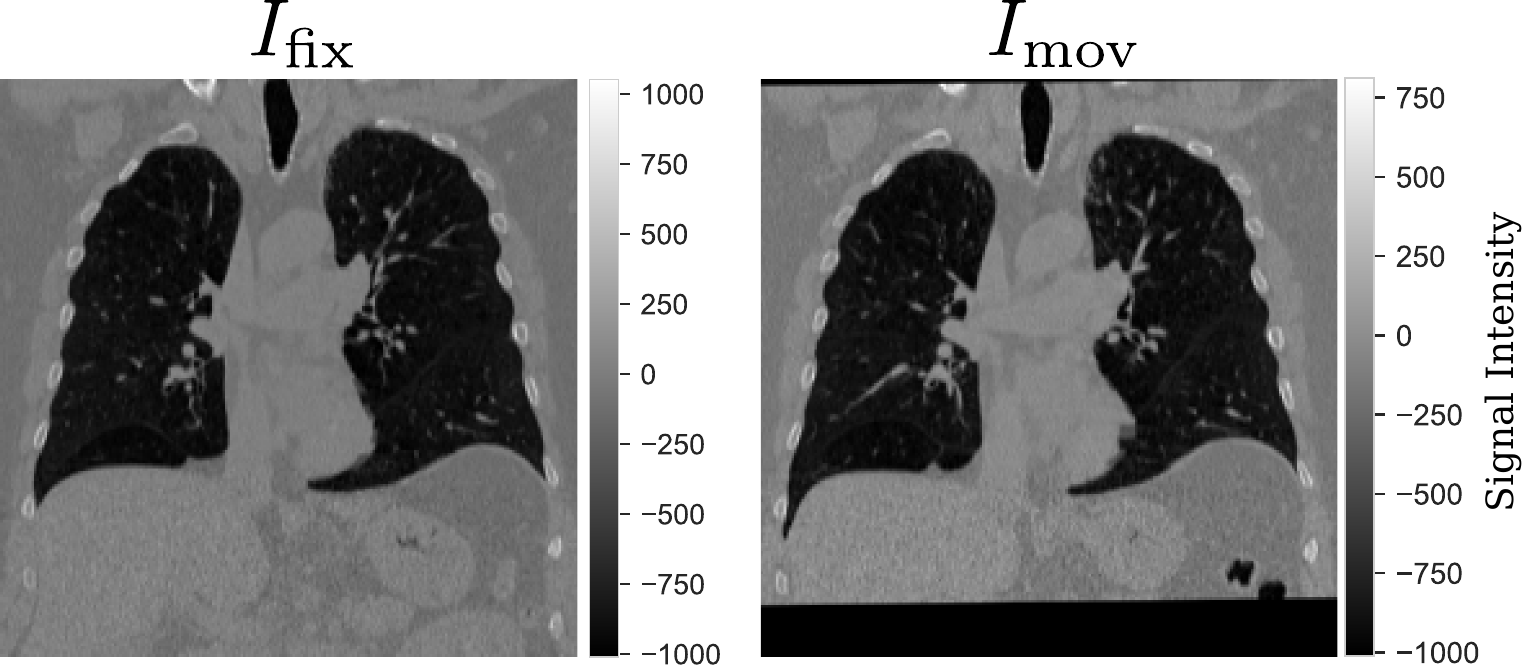}
    \caption{{An example of NLST dataset for a single subject with  $I_{\text{fix}}$ corresponding to the image captured at inspiratory phase and $I_{\text{mov}}$ corresponding to the image captured at expiratory phase.}}
    \label{fig:nlst_example}
\end{figure}

\begin{figure}[htb!]
    \centering
    \includegraphics[scale = 0.24]{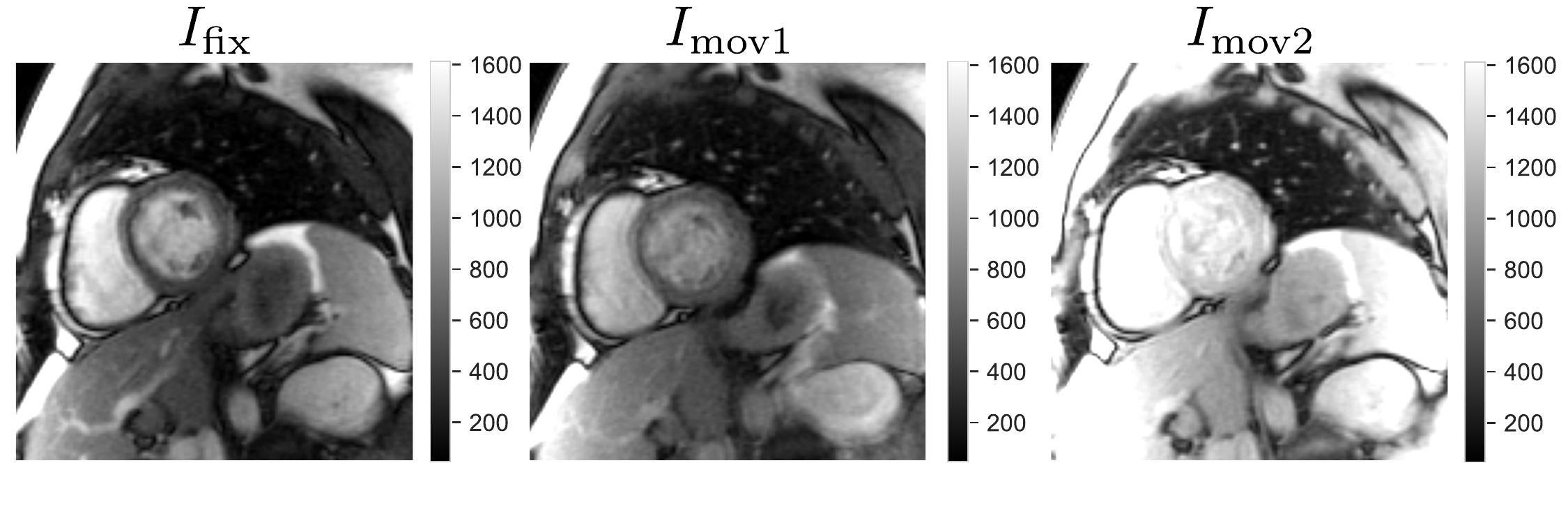}
    \caption{An example of mSASHA dataset, from left to right: $I_{\text{fix}}$, $I_\text{mov}$ (random sample 1), and $I_\text{mov}$ (random sample 2). The three images were taken from the same image series, with different acquisition time points. To emphasize the difference in both signal intensity and contrast across images in a single series, the color ranges are set to be the same for the three images.} 
    \label{fig:msasha_example}
\end{figure}

\noindent \textbf{OASIS}: The dataset contains 414 subjects, where for each subject, the normalized $T_1$-weighted scan was acquired. The subjects are split into train/validation/test with counts of $[300,30,84]$. For training, images are randomly paired using an on-the-fly data loader, while in the validation and test sets, all images are paired with the next image in a fixed order. The dataset was preprocessed with FreeSurfer and SAMSEG by \cite{hoopes2021hypermorph}, resulting in skull-stripped and bias-corrected 3D volumes with a size of $160 \times 192 \times 224$. We further resampled the images into a size of $128 \times 128 \times 128$ with intensity clipping between  $(1\%, 99\%)$ percentiles. Fig. \ref{fig:oasis_example} illustrates an example pair of OASIS images, showcasing the consistency in signal intensity and contrast.


{\noindent \textbf{NLST}: We use the public NLST dataset from the Learn2Reg challenge. The dataset consists of 210 subjects with intra-subject inhale and exhale lung CT images that are affinely pre-aligned. The subjects are split into train/validation/test with counts of $[170, 10, 30]$. Following the convention, during training, only the inhale and exhale images from the same subject are paired, with the inhale image chosen as $I_{\text{fix}}$ for all pairs. The images were preprocessed and resampled to a size of $224 \times 192 \times 224$. The keypoints and their correspondences in the lung lobe, along with the lung lobe mask, were provided by the organizers, and obtained by the corrField algorithm \mbox{\citep{heinrich2015estimating}} and nnU-Net \mbox{\citep{isensee2021nnu}} respectively. An example pair of NLST images is shown in Fig. \mbox{\ref{fig:nlst_example}}, demonstrating the large deformation required for this task.}

\noindent \textbf{mSASHA}: During an free-breathing mSASHA examination, a time series of $N =30$ real-valued 2D images, denoted by $I  = \{I_n |\  n = 1,2,\ldots,N\}$, are acquired for the same subject. In the setting of quantitative MRI, we aim to spatially align $N$ images in a single sequence $I$ into a common fixed template image $I_\text{fix}$, by individually performing $N$ pairwise registration processes over $(I_n, I_\text{fix})$ where $n = 1,2,\ldots, N$. 

The mSASHA acquisition technique \citep{chow2022improved} is a voxel-wise 3-parameter signal model based on the joint cardiac $T_1$-$T_2$ signal model:
\begin{equation}
\mathcal{S}_{}\left(T_1, T_2, A\right)=A\left\{1-\left[1-\left(1-e^{-T S / T_1}\right) e^{-T E / T_2}\right] e^{-T D / T_1}\right\},
\label{eq:msasha}
\end{equation}
where $(TS, TE, TD)$ denotes the set of three acquisition variables, and $(T_1, T_2, A)$ is the set of parameters to be estimated for each voxel coordinate of the image series. {The sequence for mSASHA consists of a reference image without magnetization preparation, a series of saturation recovery (SR) images, and a series of both SR and $T_2$-prepared images.} We encourage interested readers to refer to \cite{chow2022improved} for a more detailed explanation.

In our experiment, an in-house mSASHA dataset was used. This fully anonymized raw dataset was provided by NIH, and was considered ``non-human subject data research” by the NIH Office of Human Subjects Research''. The dataset comprises 120 subjects, with each subject having 3 slice positions, resulting in a total number of 360 slices. Each mSASHA time series consists of a fixed length of $N = 30$ images. We split \textbf{mSASHA} into train/validation/test with counts of $[84, 12, 24]$ by subjects to avoid data leakage across the three slices. Given variations in image sizes due to different acquisition conditions, we first center-cropped the images into the same size of $144 \times 144$. Subsequently, we applied intensity clipping between $(1\%, 95\%)$ percentiles to mitigate extreme signal intensities from the chest wall region. We selected the last image in the series, {\textit{i.e.,} in the $T_2$ preparation stage}, as the template $I_{\text{fix}}$, which is then a T2-weighted image with the greatest contrast between the myocardium and adjacent blood pool. An illustrative example of mSASHA images can be found in Fig. \ref{fig:msasha_example}, showing varying contrasts and non-rigid motion across frames.
\subsection{Evaluation Metrics}
{For evaluating the smoothness of the deformation field, we employ three complementary metrics based on the Jacobian of the deformation $J_{\bm{\phi}} = \nabla \bm{\phi}$: (1) the percentage of negative Jacobian determinant $|J_{\phi}|_{\leq 0}$, which quantifies the proportion of regions exhibiting folding or topology-breaking transformations, (2) the standard deviation of the log-Jacobian determinant $\text{std}(\log \vert J_{\bm{\phi}} \vert)$, which characterizes the global variation of volume changes, and (3) the magnitude of the spatial gradient of the Jacobian determinant $\operatorname{mag}(\nabla \vert J_{\bm{\phi}}\vert)$, which measures the local rate of change in volume deformation. To allow fair comparisons between methods, we meticulously adjust the regularization parameter $\lambda$ for each baseline to achieve comparable levels of smoothness of transformation. To assess registration accuracy, we utilize structural similarity metrics that are independent of optimization objectives $\mathcal{L}_{\text{sim}}$ and $\mathcal{L}_{\text{reg}}$, with dataset-specific metrics detailed in subsequent sections.}

For OASIS, two metrics, Dice score and Hausdorff distance (HD) are considered to evaluate segmentation quality after registration. Given two sets $X\subset M$ and $Y \subset M$, the Dice score, is defined to measure the overlapping of $X$ and $Y$: 
\begin{equation}
Dice(X,Y)=\frac{2|X \cap Y|}{|X|+|Y|}.
\end{equation}
Similarly, the Hausdorff distance of two aforementioned sets $X$ and $Y$ is given by:
\begin{equation}
HD(X, Y):=\max \left\{\sup _{x \in X} d(x, Y), \sup _{y \in Y} d(X, y)\right\},
\end{equation}
where $d(\cdot, \cdot)$ is a metric (2-norm in this work) on $M$ and $d(x, Y):=\inf _{y \in Y} d(x, y)$. As a remark, in this work, we consider the average across all segmentation labels to calculate the Dice score and HD in \textbf{OASIS} instead of only considering the major regions.

{For the NLST dataset, we evaluate the registration performance using both lung lobe mask overlap and Target Registration Error (TRE) of the keypoints. For a quantitative evaluation of local registration accuracy, TRE is calculated based on the provided keypoint pairs. Given a set of corresponding keypoint pairs $\{(\bm{p}_i, \bm{q}_i)\}_{i=1}^L$ where $\bm{p}_i$ represents a keypoint in the moving image and $\bm{q}_i$ its corresponding point in the fixed image, TRE is defined as:}
\begin{equation}
    \text{TRE} = \frac{1}{L}\sum_{i=1}^L \|\bm{\phi}(\bm{p}_i) - \bm{q}_i\|_2,
\end{equation}
{where $\|\cdot\|_2$ denotes the Euclidean distance.}

Furthermore, we also evaluate two more independent metrics for the mSASHA dataset proposed by \cite{huizinga2016pca} isolated from training. The metrics are based on the principal component analysis (PCA) of images. Assume $M \in \mathbb {R}^{d_xd_y \times N}$ is the matrix representation of $I$, where a row of $M$ represents a coordinate in the image space. The correlation matrix of $M$ is then calculated by:
\begin{equation}
K=\frac{1}{d_xd_y-1} \Sigma^{-1}(M-\overline{M})^{\intercal}(M-\overline{M}) \Sigma^{-1},
\end{equation}
where $\Sigma$ is a diagonal matrix representing the standard deviation of each column, and $\overline{M}$ denotes the column-wise mean for each column entry. Since an ideal qMRI model assumes a voxel-wise tissue alignment, the actual underlying dimension of $K$ can be characterized by a low-dimensional (linear) subspace driven by the signal model. In the mSASHA signal model, the dimension of such a subspace is assumed to be four according to Eq. \ref{eq:msasha}, determined by the number of parameters to be estimated. With the fact that the trace of $K$, $\text{tr}(K)$ is a constant, two PCA-based metrics were proposed as follows:

\begin{align}
\mathcal{D}_{\text{PCA1}}&=\sum_{i=1}^N \sigma_{i}-\sum_{j=1}^L \sigma_j=\text{tr}(K)-\sum_{j=1}^L\sigma_j ,\\
\mathcal{D}_{\text{PCA2}}&=\sum_{j=1}^N j \sigma_j,
\end{align}
{where $\sigma_i$ denotes the $i$-th largest eigenvalue of $K$}. Both metrics were designed to penalize a long-tail distribution of the spectrum of $K$, and $L$ is a hyperparameter regarding the number of parameters of the signal model. For $\mathcal{D}_{\text{PCA1}}$, an ideal scenario would involve all images perfectly aligning with tissue anatomy and the signal model, resulting in a value of $0$. Meanwhile, the interpretation of $\mathcal{D}_{\text{PCA2}}$ further emphasizes the tail of the eigenvalues, thus enlarging the gaps across experiments. 

To narrow the analysis to the region of interest to the heart region, the calculation is confined to this area by cropping the resulting images before computing the metric. This constraint ensures that the evaluation is focused on the relevant anatomical structures.

\begin{table*}[htb!]
\centering
\tiny
\caption{Quantitative comparison of different registration methods with $100\%$ data availability. For each evaluation metric except Par. (number of parameters), Mem. (VRAM consumption), and $T$ (training/inference time), we report mean$\pm$std. VRAM consumption indicates the peak GPU memory consumption during training. The training/inference time is profiled for a batch on GPU (excluding data loading time), except for elastix where CPU time is reported. For LapIRN, metrics are reported at level 3. The best performance is shown in bold and marked with $*$ if there is a statistically significant difference ($p < 0.05$) from the second best by a two-sided Wilcoxon signed-rank test.}
\label{tab:summary}
\begin{tabular}{>{\arraybackslash}p{0.86cm}||>{\centering\arraybackslash}p{0.6cm}>{\centering\arraybackslash}p{1.8cm}>{\centering\arraybackslash}p{1.8cm}|>{\centering\arraybackslash}p{1.4cm}>{\centering\arraybackslash}p{1.4cm}>{\centering\arraybackslash}p{1.4cm}|>{\centering\arraybackslash}p{0.5cm}>{\centering\arraybackslash}p{0.5cm}>{\centering\arraybackslash}p{0.8cm}}
\toprule
Dataset & Method & Dice & HD(mm) & $|J_{\bm{\phi}}|_{\leq 0}(\%)$ & std$(\log|J_{\bm{\phi}}|)$ &$\operatorname{mag}(\nabla \vert J_{\bm{\phi}}\vert)$& Par.(K) & Mem.(G) & $T$ (s) \\
\cmidrule{3-10}
\multirow{8}{*}{OASIS} 
& Affine & $0.563\pm0.063$ & $8.56\pm1.19$ & - & -&- &  - & - & - \\
& elastix & $0.696\pm0.040$ & $3.89\pm0.58$ & $0.003\pm0.002$ & $0.254\pm0.045$ & $0.015 \pm 0.002$ &- & - & 17.45 \\
& VM & $0.729\pm0.030$ & $3.61\pm0.49$ & $0.016\pm0.031$ & $0.295\pm0.115$ & $0.029\pm 0.003$& 320 & 1.57 & 0.17/0.14 \\
& L2O & $0.733\pm0.032$ & $3.62\pm1.01$ & $0.021\pm0.028$ & $0.411\pm0.132$ &$0.035\pm 0.003$& 343 & 20.52 & 1.19/0.86 \\
& GraDIRN & $0.739\pm0.027$ & $3.54\pm0.45$ & $0.012\pm0.003$ & $0.206\pm0.065$ & $0.022\pm 0.002$&269 & 16.13 & 0.57/0.25 \\
& LapIRN & $0.751\pm0.026$ & $3.50\pm0.44$ & $0.010\pm0.002$ & $0.260\pm0.051$ & $0.034\pm 0.002$&923 & 3.78 & 0.16/0.10 \\
& RCVM & $0.753\pm0.025$ & $\bm{3.46\pm0.43}^*$ & $0.002\pm0.001$ & $0.189\pm0.043$ & $0.027\pm 0.003$& 1920 & 10.20 & 0.64/0.54 \\
& RIIR  & $\bm{0.756\pm0.025}^*$ & $3.48\pm0.41$ & $0.011\pm0.009$ & $0.264\pm0.073$ & $0.029\pm 0.003$&436 & 11.82 & 0.55/0.27 \\
\midrule

Dataset & Method & TRE(mm) & Dice & $|J_{\bm{\phi}}|_{\leq 0}(\%)$ & std$(\log|J_{\bm{\phi}}|)$&$\operatorname{mag}(\nabla \vert J_{\bm{\phi}}\vert)$ & Par.(K) & Mem.(G) & $T$(s) \\
\cmidrule{3-10}
\multirow{8}{*}{NLST} 
& Affine & $8.43\pm3.97$ & $0.873\pm0.041$ & - & - & - & - & - & - \\
& elastix & $5.01\pm 2.92$ & $0.946\pm 0.007$ & $0.000\pm0.000$ & $0.160\pm0.031$ & $0.011\pm 0.003$& - & - & 14.21 \\
& VM & $3.99\pm2.48$ & $0.957\pm0.012$ & $0.120\pm0.162$ & $0.677\pm0.399$ & $0.057\pm 0.005$& 320 & 0.93 & 0.18/0.15 \\
& L2O & $2.51\pm2.96$ & $0.961\pm0.019$ & $0.451\pm0.363$ & $1.325\pm0.554$ & $0.068\pm 0.007$& 343 & 11.81 & 0.83/0.52 \\
& GraDIRN & $2.89\pm2.49$ & $0.960\pm0.020$ & $0.114\pm0.068$ & $0.454\pm0.269$ & $0.058\pm 0.005$&279 & 9.32 & 0.37/0.15 \\
& LapIRN & $2.44\pm1.60$ & $0.967\pm0.006$ & $0.073\pm0.065$ & $0.555\pm0.240$ &$0.058\pm 0.004$ &923 & 2.18 & 0.14/0.05 \\
& RCVM & $2.32\pm1.35$ & $\bm{0.968\pm0.005}^*$ & $0.229\pm0.278$ & $0.913\pm0.511$ & $0.063\pm 0.005$& 1920 & 5.89 & 0.38/0.27 \\
& RIIR  & $\bm{2.21\pm1.71}^*$ & $0.966\pm0.012$ & $0.108\pm0.259$ & $0.568\pm0.481$ & $0.048\pm 0.004$ &436 & 7.12 & 0.41/0.18 \\
\midrule
Dataset & Method & $\mathcal{D}_{\text{PCA1}}$ & $\mathcal{D}_{\text{PCA2}}$ & $|J_{\bm{\phi}}|_{\leq 0}(\%)$ & std$(\log|J_{\bm{\phi}}|)$&$\operatorname{mag}(\nabla \vert J_{\bm{\phi}}\vert)$ & Par.(K) & Mem.(G) & $T$(s) \\
\cmidrule{3-10}
\multirow{8}{*}{mSASHA} 
& Raw & $1.28\pm0.88$ & $46.70\pm10.40$ & - & - & - & - & - & - \\
& elastix & $0.40\pm 0.38$ & $35.81\pm 4.72$ & $0.162\pm0.005$ & $0.551 \pm 0.685$ & $0.022\pm 0.009$ & - & - & 2.46 \\
& VM & $0.46\pm0.42$ & $36.49\pm5.13$ & $0.002\pm0.001$ & $0.210\pm0.082$ & $0.053\pm0.012$ &79 & 0.06 & 0.05/0.01 \\
& GraDIRN & $0.39\pm0.38$ & $35.85\pm4.51$ & $0.002\pm0.002$ & $0.156\pm0.043$ & $0.040\pm0.012$ & 89 & 0.21 & 0.09/0.04 \\
& LapIRN & $0.34\pm0.32$ & $35.11\pm4.09$ & $0.006\pm0.002$ & $0.248\pm0.128$ & $0.054\pm0.015$ & 309 & 0.09 & 0.04/0.01 \\
& RCVM & $\bm{0.32\pm0.30}^*$ & $\bm{34.79\pm3.83}^*$ & $0.005\pm0.002$ & $0.275\pm0.081$ & $0.073\pm0.019$ &589 & 0.13 & 0.07/0.02 \\
& RIIR & $0.36\pm0.36$ & $35.46\pm4.41$ & $0.007\pm0.002$ & $0.312\pm0.042$ & $0.039\pm0.008$ & 148 & 0.24 & 0.13/0.05 \\

\bottomrule
\end{tabular}
\end{table*}

\subsection{Experimental Settings}
We here summarize the main experiments for evaluation and further ablation experiments for RIIR. For all experiments, the main workflow is to register the image series $I$ of length $N$ in a pairwise manner: that is, we first choose a template $I_\text{fix}$, and then perform $N$ registrations. When $N = 2$, the registration process simplifies to straightforward pairwise registration.
 
\begin{figure*}[htb!]
    \centering
    \includegraphics[scale = 0.42]{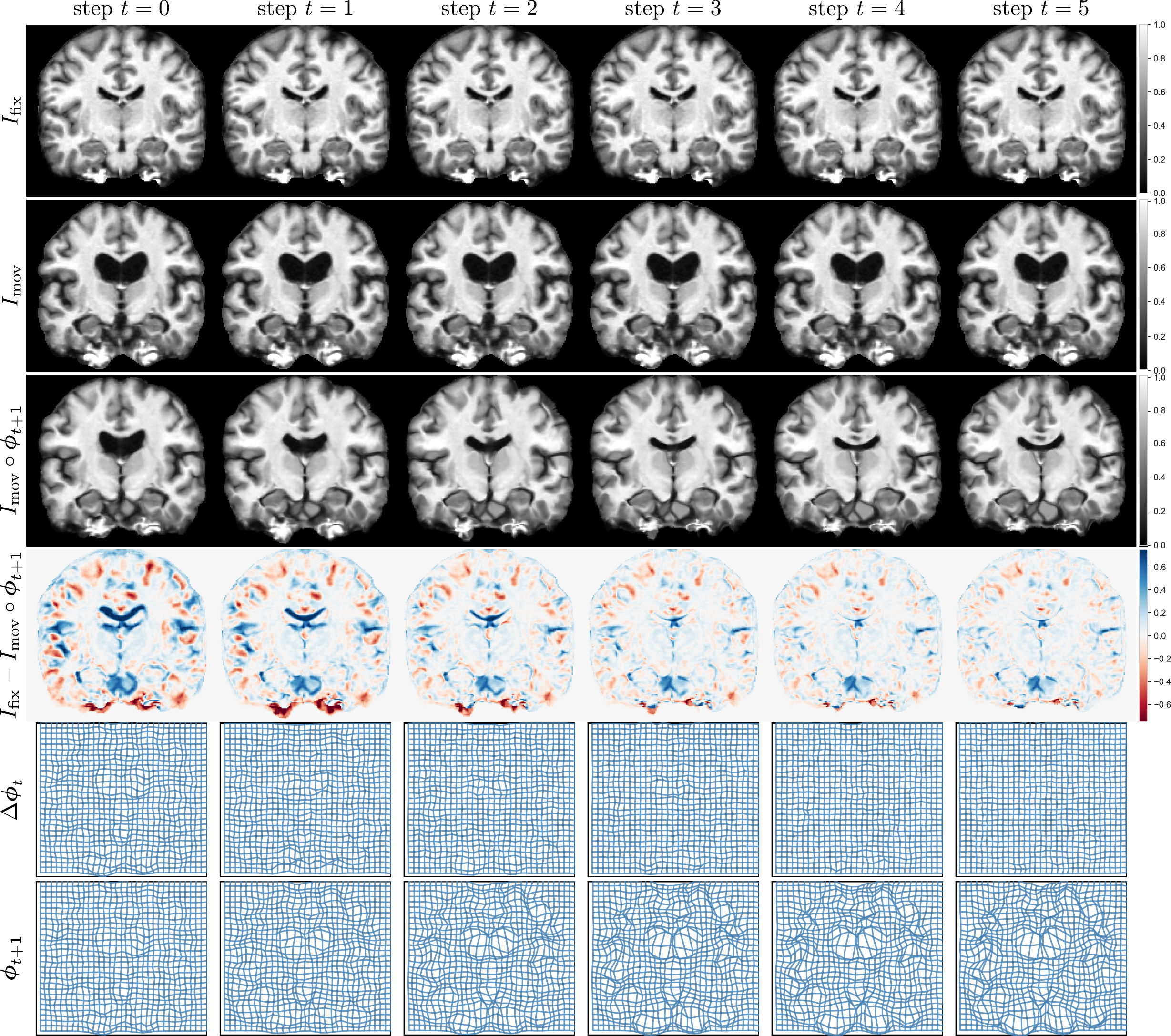}
    \caption{{A visualization of RIIR inference on OASIS test split, visualized with a 2D slice and in-plane deformation. The inference step was set to $6$ in both training and inference. All images in the same row were plotted using the same color range for better consistency.}}
    \label{fig:riir_vis_oasis}
\end{figure*}
\noindent\textbf{Experiment 1: Comparison Study with Varying Data Availability}

\noindent We introduce five data-availability scenarios to evaluate the robustness of the models when data availability is limited, on both datasets, which often happens in both research settings and clinical practices as the number of subjects is heavily limited. The training data availability settings in this study were set to $[5\%, 10\%, 25\%, 50\%, 100\%]$ for {all datasets}. It is worth noticing that for limited data availability scenarios, the data used for training remained the same for all models in consideration, and the leave-out test split remained unchanged for all scenarios.

\noindent\textbf{Experiment 2: Inclusion of Hidden States}

\noindent Unlike most related works utilizing the original RIM framework \citep{lonning2019recurrent,sabidussi2021recurrent} where two levels of hidden states are considered, we explored the impact of modifying or even turning off hidden states. In our implementation of convolutional GRU, at most two levels of hidden states $\boldsymbol{h}_t^1$ and $\boldsymbol{h}_t^2$ are considered, following recent works using RIM \citep{lonning2019recurrent,sabidussi2021recurrent,sabidussi2023dtirim}. Both hidden states were configured with 32 channels in their corresponding convolutional GRU layers. All experiments in this ablation study were performed in the validation split. {We present the results for OASIS and NLST as they represent distinct imaging modalities (brain MRI and lung CT) and registration scenarios}.

\noindent\textbf{Experiment {3}: Inclusion of Gradient of Inner Loss $\nabla_{\bm{\phi}_t} \mathcal{L}_{\text{inner}}$ as RIIR Input}

\noindent We performed an experimental study on the input composition for RIIR. As shown in Eq. \ref{update riir}, the goal was to study the data efficiency and the registration performance by incorporating the gradient of $\mathcal{L}_{\text{inner}}$ in RIIR. We could achieve ablation by changing the input of $g_\theta$. A comparison with other input modeling strategies seen in \cite{qiu2022embedding} was proposed against the gradient-based input for $g_\theta$. Depending on whether the moving image is deformed (explicit) or not (implicit), {as well as the original RIM formulation, we ended up with four input compositions}:
\begin{enumerate}
    
    \item Implicit Input without $\nabla \mathcal{L}_{\text{inner}}$: $[\bm{\phi}_t, I_\text{mov}, I_\text{fix}]$;
    \item Explicit Input without $\nabla \mathcal{L}_{\text{inner}}$: $[\bm{\phi}_t, I_\text{mov} \circ \bm{\phi}_t, I_\text{fix}]$;
    \item {RIM Input: $[\bm{\phi}_t, \nabla_{\bm{\phi}_t}\mathcal{L}_{\text{inner}}]$;}
    \item {RIIR Input: $[\bm{\phi}_t, \nabla_{\bm{\phi}_t}\mathcal{L}_{\text{inner}}, I_\text{mov} \circ \bm{\phi}_t, I_\text{fix}]$.}
\end{enumerate}
This study aimed to provide information on the impact of different input compositions on the efficiency of RIIR when data availability varies. We conducted the experiment with {two data availability choices ($[5\%,100\%]$) to examine the data efficiency and other potential influences induced by the gradient input. }

\noindent \textbf{{Experiment 4: Regularization Analysis}}

\noindent {It is known that the regularization metric and weight influence the registration performance. In this experiment, we want to investigate the change in the evaluation metrics. For comparison, we used diffusion, curvature, and elastic regularization functions. For elastic regularization, we set the elastic parameters to be $[\lambda_e=540.8, \mu=22.5]$ for OASIS and $[\lambda_e=45.33, \mu=8]$ for NLST according to the literature} \citep{kumaresan1996importance,lai2000effects,reithmeir2024learning}.

\noindent \textbf{Experiment 5: RIIR Architecture Ablation}
\noindent Since RIIR is the first attempt to formulate and implement the RIM framework for medical image registration, we performed an ablation study on the RIIR network architecture for the number of evaluation steps $T$.

\subsection{Baseline Methods and Implementation Details}
{We compared our proposed method to various registration methods that are closely related to our interest. We use the same choice of similarity and regularization functions for all deep learning-based methods in the comparative study for fair comparison, unless otherwise indicated. We used MSE loss for OASIS, NCC with window size $w=5$ for NLST, and NMI with $n=32$ bins for mSASHA. We used diffusion regularization for all three datasets in the comparative study. The methods and applicable hyperparameters are described as follows:}

\begin{itemize}
    \item {Elastix \mbox{\citep{klein2009elastix}}: An iterative optimization-based registration toolbox. Specifically, we used ITK-Elastix \mbox{\citep{ntatsis2023itk}} in Python. Three resolution levels with third-order B-spline transformation and a grid spacing of four were applied for all datasets.}
    \item VoxelMorph \citep{balakrishnan2019voxelmorph}: {We used $\lambda = 0.02$ for OASIS as in the original paper, $\lambda= 0.15$ for NLST, and $\lambda = 0.3$ for mSASHA. The channels used in each downsampling encoder block were $\left[ 16,16,32,32,32 \right]$. Two layers of activation with channels $\left[16, 16\right]$ after the decoder was used.}
    \item GraDIRN \citep{qiu2022embedding}: {A multi-resolution multi-step deep learning method that uses explicit similarity loss gradient and dense CNN to produce incremental updates. We followed the original implementation with 3 resolutions and 3 steps per resolution and use the last-step output for loss calculation. The training losses were set to the same as VoxelMorph, with weight parameter $\lambda = 0.015$ for OASIS, $\lambda= 0.125$ for NLST, and $\lambda = 0.25$ for mSASHA.}
    \item {LapIRN \mbox{\citep{mok2020large}}: A multi-resolution deep learning method with Laplacian pyramid networks. We followed the original implementation of the displacement version with three-stage training and used multi-level NCC in the paper as similarity loss for OASIS, with $\lambda = 0.6$. For NLST and mSASHA, we followed the same loss and weight as for VoxelMorph. }
    \item  {Learn-to-Optimise (L2O) \mbox{\citep{falta2022learning}}: A multi-step deep learning method that uses additional input modalities, including 3D MIND features \mbox{\citep{heinrich2012mind}} and sampled coordinates for 3D datasets. Constrained by the inherent 3D nature of MIND features, we evaluated L2O on the OASIS and NLST datasets. The original paper uses keypoint supervision only; therefore, we replaced the loss function by the same unsupervised loss as used by other baseline methods while keeping the uniform weight for each step, with $\lambda = 0.03$ for OASIS and $\lambda = 0.225$ for NLST.}
    \item {Recursive-cascaded VoxelMorph (RCVM)} \citep{zhao2019recursive}:{A single-resolution iterative deep learning method that uses cascaded U-Nets and composition of the deformation field to generate the final output. We used the same VoxelMorph backbone and loss functions with $\lambda = 0.015$ for OASIS, $\lambda = 0.25$ for mSASHA, and $\lambda= 0.125$ for NLST.}
\end{itemize}

We implemented the RIIR in the following settings for experiment 1: The backbone network is the same as the VoxelMorph in baseline, with additional ConvGRU in the second level with 32 channels. We used the inference steps $T = 6$ and exponential weighting for $w_t = 10^{\frac{t-1}{T-1}}$. {To ensure a similar level of smoothness, the trade-off parameter for RIIR was set to $\lambda = 0.0125$ for OASIS, $\lambda = 0.125$ for NLST, and $\lambda = 0.25$ for mSASHA. The optimizer of all methods remained the same using Adam \mbox{\citep{adampaper}} with $\beta_1 = 0.9$ and $\beta_2 = 0.999$. The initial learning rate was set to $1\times 10^{-4}$ for all models. For all experiments, the maximum epochs was set to 100 epochs with early stopping if the evaluation metrics does not improve for $10$ epochs.} The experiments were performed on an NVIDIA RTX 4090 GPU with a VRAM of 24 GB. {The source codes for RIIR, with the implementation of the baseline models and data processing are publicly available\footnote{\url{gitlab.tudelft.nl/ai4medicalimaging/riir-public}}}.

\section{Results}

\subsection{Experiment 1: Comparison Study with Varying Data Availability}

\begin{figure}[htb!]
    \centering  
    \includegraphics[width=0.6\textwidth]{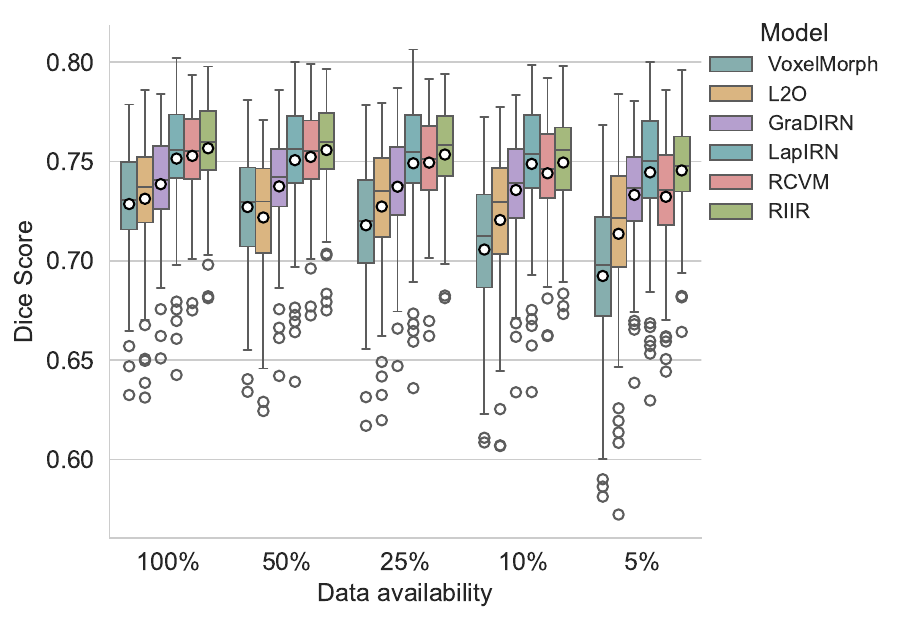}

    \caption{{Results of Experiment 1 with boxplots for Dice score on OASIS. The circle denotes the mean of the metric of interest. The segmentation metric Dice is calculated for all 35 segmentation labels and post-processed by taking the average. }}
    \label{fig:ex1_oasis}
    
\end{figure}

 An illustrative visualization of RIIR inference on an example test data, can be found in Fig. \ref{fig:riir_vis_oasis}. The results for \textbf{OASIS} are presented in Fig \ref{fig:ex1_oasis}, as well as . {LapIRN, leveraging its multi-resolution architecture, also demonstrates robust performance}. It is evident that RIIR outperforms most deep learning-based baselines when data availability is severely limited and maintains consistent performance across various data availability scenarios, showcasing its data efficiency and accuracy. 
 
\begin{figure}[htb!]
    \centering  
    \includegraphics[width=0.6\textwidth]{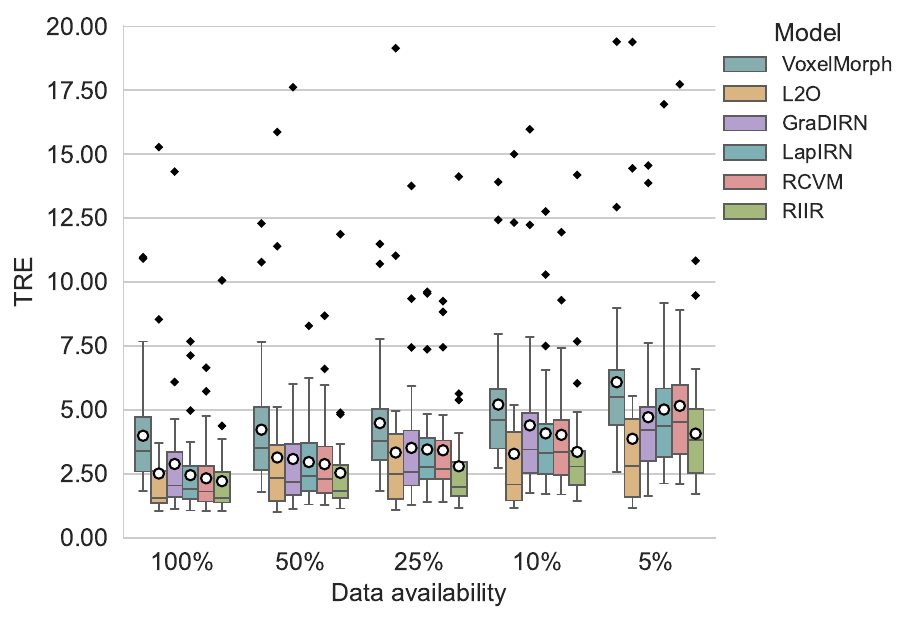}
    \caption{{Results of Experiment 1 with boxplots for TRE on NLST. The circle denotes the mean of the metric of interest. The TRE is calculated based on the anatomically meaningful keypoint pairs provided by corrField algorithm within lung lobes. For visualization, outliers over a TRE of 20mm were excluded but used for statistical calculation.}}
    \label{fig:ex1_nlst}
 \end{figure}
 
 \begin{figure}[htb!]
     \centering  
     \includegraphics[width=0.6\textwidth]{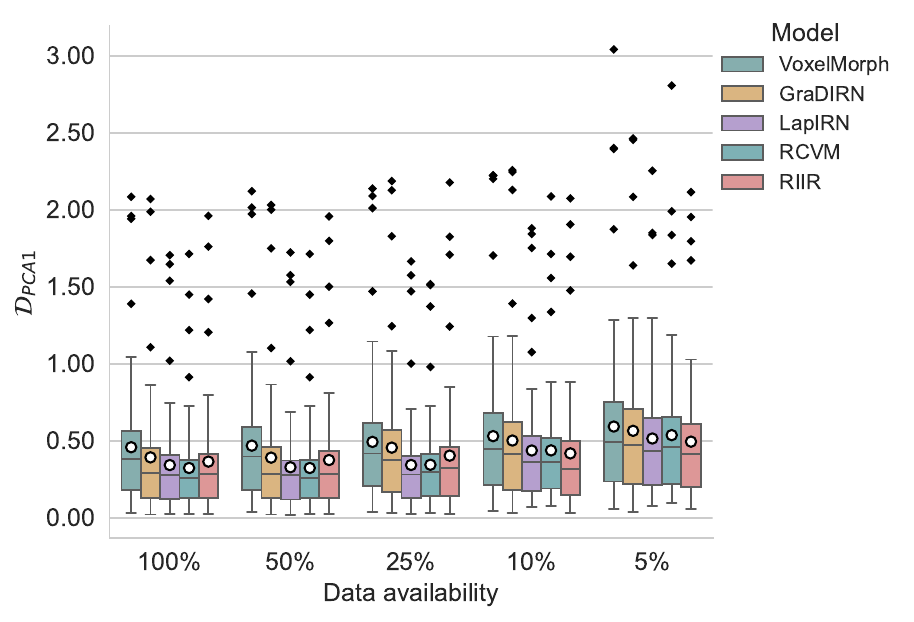}
     \caption{{Results of Experiment 1 with boxplots for $D_{pca1}$ on mSASHA. The circle denotes the mean of the metric of interest. The group-wise metric $D_{\text{PCA1}}$ was calculated based on further center-cropping at a ratio of $70\%$ on the warped images.}}
     \label{fig:ex1_msasha}
   \end{figure}

\begin{figure*}[hb!]
    \centering
    \includegraphics[scale = 0.35]{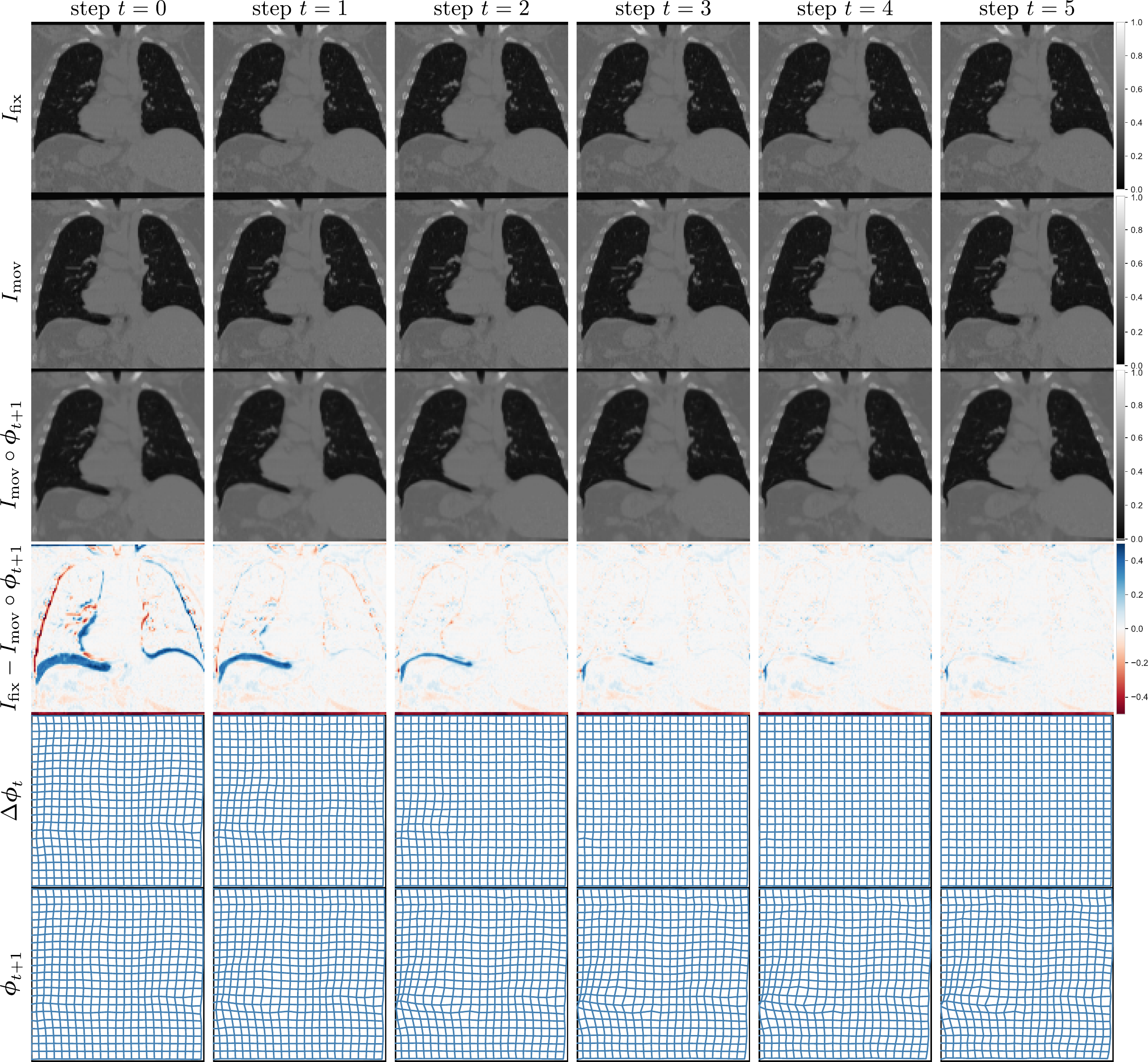}
    \caption{{A visualization of RIIR inference on NLST test split, visualized with a 2D slice and in-plane deformation. The inference step was set to $6$ in both training and inference. All images in the same row were plotted using the same color range for better consistency.}}
    \label{fig:riir_vis_nlst}
\end{figure*}

{The results for \textbf{NLST} are shown in Fig. {\ref{fig:ex1_nlst}}.  An illustrative visualization of RIIR inference on an inhale-exhale lung CT pair, can be found in Fig. {\ref{fig:riir_vis_nlst}}. The registration of lung CT presents unique challenges due to the large deformation between respiratory phases. RIIR demonstrates superior performance in capturing these large deformations, achieving lower TRE while maintaining anatomically plausible transformations.}

\begin{figure}[htb!]
    \centering
    \includegraphics[scale = 0.75]{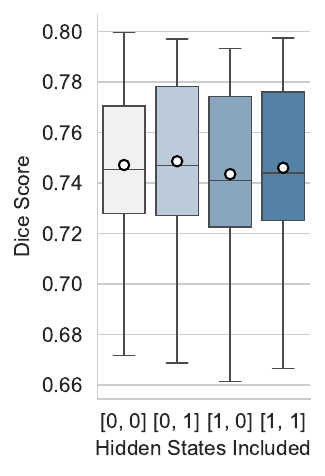}
    \includegraphics[scale = 0.75]{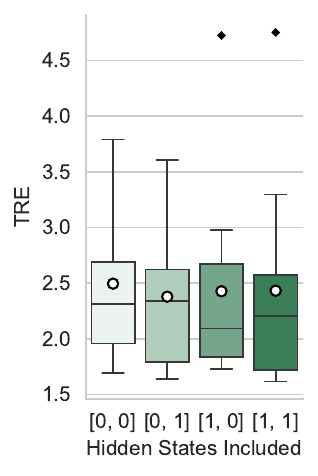}
    \caption{{Results of Experiment 2 evaluated on OASIS validation set (left) regarding Dice score and NLST validation set (right) regarding TRE. Here, for example, $[0,0]$ denotes the case that no hidden states are considered, and $[1,1]$ denotes both hidden states were considered in the pipeline. Two-sided Wilcoxon tests were conducted for $[0,1]$ against other settings with statistical significance $(p < 0.05)$, except for $[0,0]$ $(p=0.076)$ in OASIS dataset.}}
    \label{fig:ex4_box_dice_nlst}
\end{figure}

\begin{figure*}[htb!]
    \centering
    \includegraphics[scale = 0.35]{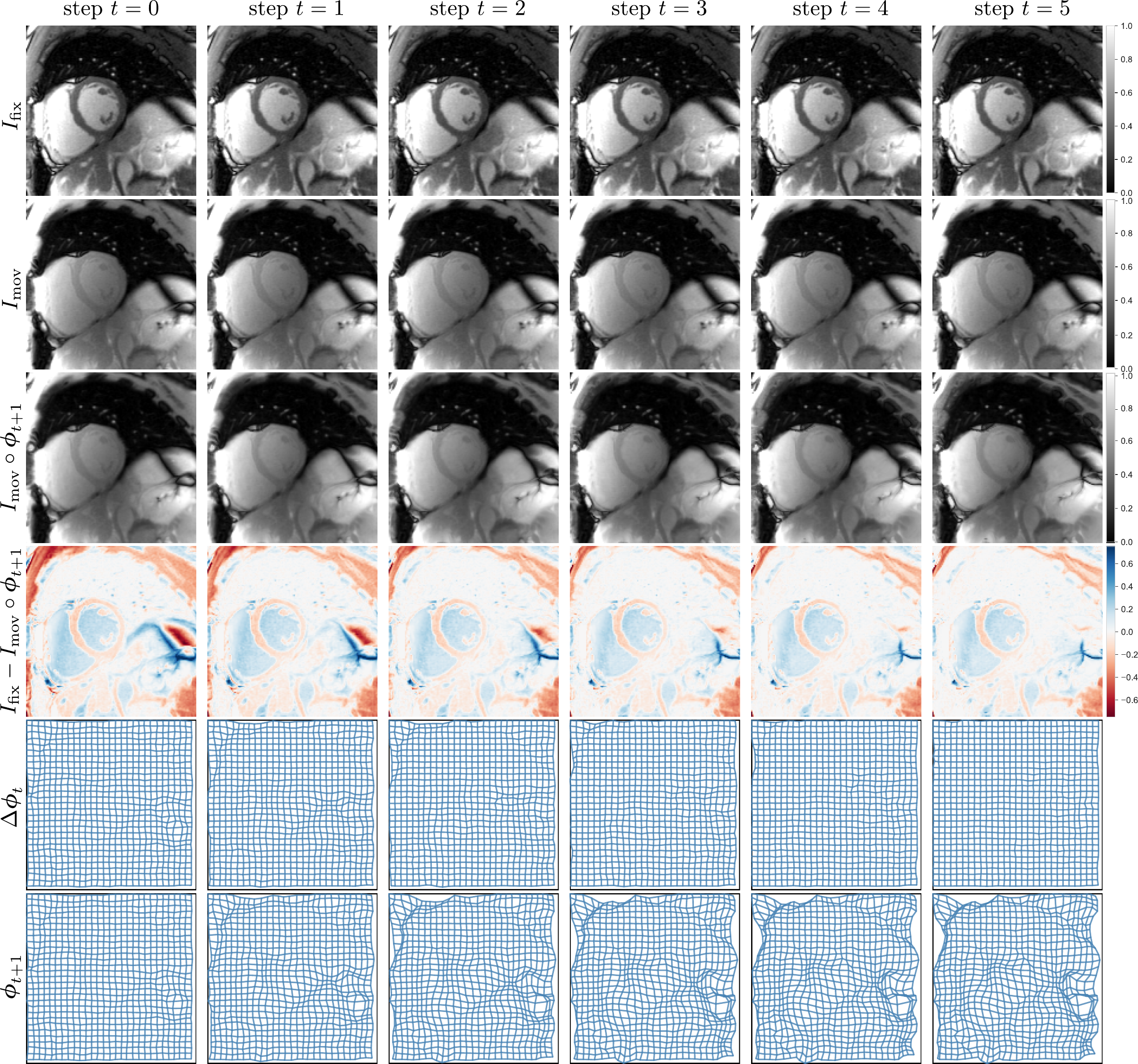}
    \caption{{A visualization of RIIR inference on \textbf{mSASHA} test split. The inference step was set to $6$ in both training and inference. All images in the same row were plotted using the same color range for better consistency. The deformation field was initialized as an identity mapping.}}
    \label{fig:riir_vis_msasha}
\end{figure*}

The results of this experiment on mSASHA are shown in Fig. \ref{fig:ex1_msasha} using a composition of boxplots. {Both LapIRN and RCVM achieved superior performance in group-wise registration, with RCVM showing slightly better results in terms of $\mathcal{D}_{\text{PCA1}}$. Our proposed RIIR demonstrated comparable performance levels in terms of $\mathcal{D}_{\text{PCA1}}$}. The qualitative visualization of RIIR inference on mSASHA test split is shown in Fig. \ref{fig:riir_vis_msasha}.


{For a comprehensive quantitative comparison, Table {\ref{tab:summary}} presents detailed statistics across all datasets under full data availability, including performance metrics, model parameters, memory consumption, and computational time.}





    

\subsection{Experiment 2: Inclusion of Hidden States}
 The architectural settings were kept the same as in the aforementioned experiments, and the results are shown in Fig. \ref{fig:ex4_box_dice_nlst}. {Although improved performance using hidden states ([0, 1]) over no hidden states is only observed in the NLST dataset, we empirically noticed that training was more stable when hidden states were enabled.} {It is also notable that the inclusion of hidden states in the pipeline does not incur significant computational overhead}. {For the OASIS dataset, the addition of hidden states only increases VRAM consumption by approximately 800MB, while maintaining the model's computational speed}. 
\begin{table}[h!]
    \centering
    \small
    \caption{{Comparison of Explicit Input and RIIR under 5\% data availability in Experiment 3. For each metric, we report mean$\pm$std.}}
    \label{tab:5percent}
    \begin{tabular}{l||ccc}
    \toprule
    Dataset & $|J_{\phi}|_{\leq 0}(\%)$ & std$(\log|J_{\phi}|)$ & $\operatorname{mag}(\nabla \vert J_{\phi}\vert)$ \\
    \midrule
    \multicolumn{4}{l}{\textbf{OASIS}} \\
    Explicit & $0.014\pm0.015$ & $0.299\pm0.133$ & $0.035\pm0.003$ \\
    RIIR & $0.019\pm0.012$ & $0.350\pm0.088$ & $0.034\pm0.003$ \\
    \midrule
    \multicolumn{4}{l}{\textbf{NLST}} \\
    Explicit & $0.092\pm0.108$ & $0.575\pm0.372$ & $0.057\pm0.003$ \\
    RIIR & $0.085\pm0.084$ & $0.581\pm0.309$ & $0.055\pm0.004$ \\
    \bottomrule
    \end{tabular}
    \end{table}

\subsection{Experiment 3: Inclusion of Inner Loss Gradient as RIIR Input}
\begin{figure}[htb!]
    \centering
    \includegraphics[scale = 0.65]{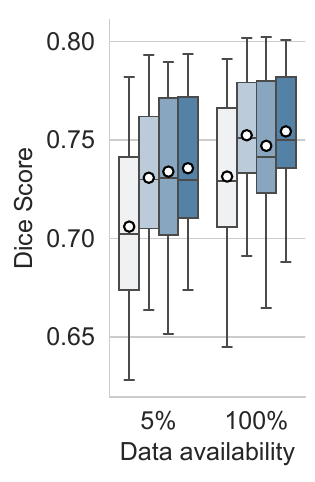}
    \includegraphics[scale = 0.65]{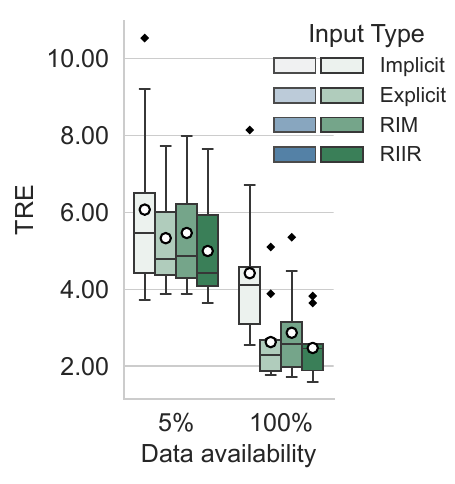}
    \caption{{Results of Experiment 3 evaluated on OASIS validation set (left) and NLST validation set (right). For OASIS with 5\% data availability, RIIR input shows significance over all types except RIM ($p=0.81$). At 100\%, significance remains except for Explicit input ($p=0.17$). For NLST with 5\% data availability, RIIR input shows significance over all types. At 100\%, significance remains except for Explicit input ($p=0.27$).}}
    \label{fig:ex3_combined}
\end{figure}

The results are shown in Fig. \ref{fig:ex3_combined}. It can be observed that the network struggled if the warped image $I_{\text{mov}} \circ \bm{\phi}$ is only implicitly fed to the RIIR cell, \textit{i.e.,} there's an additional cost to learn the transformation. {When training data is abundant (100\% availability), RIIR and Explicit input achieve comparable performance, as evidenced by the non-significant statistical differences ($p=0.17$ for OASIS and $p=0.27$ for NLST). This suggests that with sufficient training data, both approaches can effectively learn the registration task, though RIIR maintains its advantage over other input types. However, RIIR input demonstrates more robust performance in data-limited scenarios, particularly when only 5\% of the training data is available. We also found that for both datasets when data availability is $5\%$, the deformation-related metrics have smaller variation for RIIR input compared with Explicit inputs as shown in Table {\ref{tab:5percent}}.}

\subsection{Experiment 4: Regularization Analysis}
\begin{figure}[htb!]
    \centering
    \includegraphics[scale = 0.55]{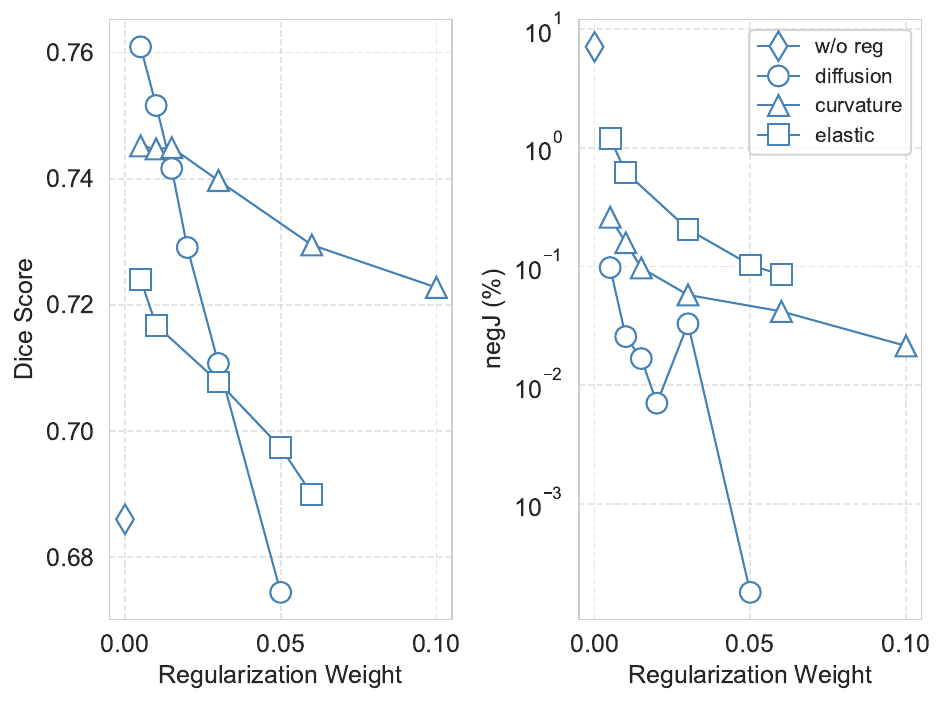}
    \caption{{Results of regularization analysis on OASIS validation set. Different regularization weights are compared in terms of Dice score. Here w/o reg denotes the results without regularization and negJ denotes the percentage of negative Jacobian determinant.}}
    \label{fig:reg_oasis}
\end{figure}

\begin{figure}[htb!]
    \centering
    \includegraphics[scale = 0.55]{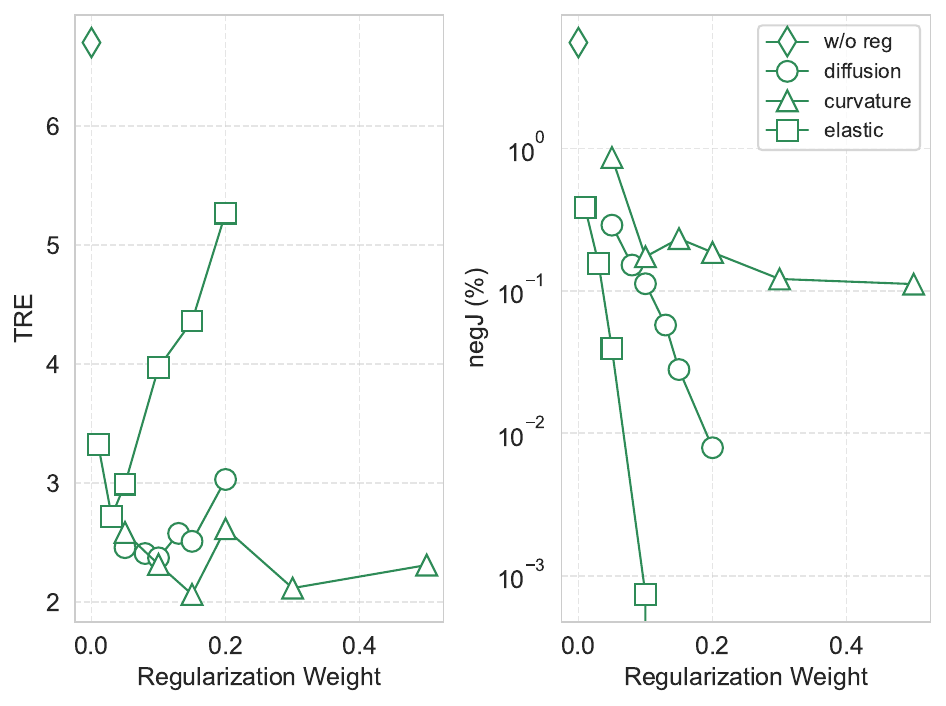}
    \caption{{Results of regularization analysis on NLST validation set. Different regularization weights are compared in terms of TRE. Here w/o reg denotes the results without regularization and negJ denotes the percentage of negative Jacobian determinant.}}
    \label{fig:reg_nlst}
\end{figure}

{The results of regularization analysis on OASIS and NLST datasets are shown in Fig. {\ref{fig:reg_oasis}} and Fig. {\ref{fig:reg_nlst}}, respectively. For OASIS, diffusion regularization achieves better Dice scores compared to affine registration across all weights. For NLST, while {all three regularization terms} improve the TRE over affine registration, curvature regularization exhibits a higher percentage of negative Jacobian determinants. {For elastic regularization, it is worth noting that different parameter settings for $\lambda_e$ and $\mu$ were used across datasets to reflect tissue differences, which may lead to varying regularization effects.}}

\subsection{Experiment 5: RIIR Architecture Ablation}
\begin{figure}[htb!]
    \centering
    \includegraphics[width=0.6\linewidth]{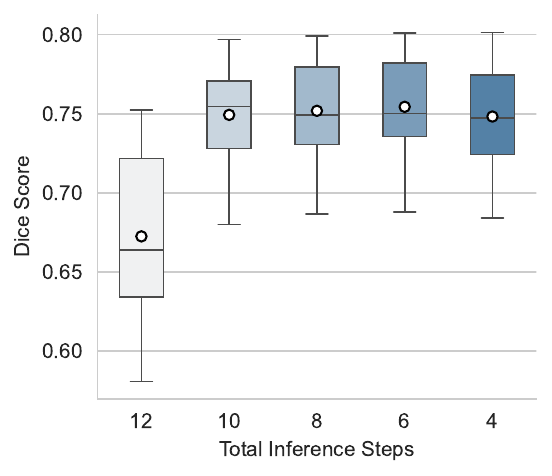}
    \caption{Results of Experiment 5. For the ablation study on network steps on OASIS, two-sided Wilcoxon tests suggest significant difference is found for $t = 6$ against all other scenarios.}
    \label{fig:ex5}
\end{figure}

Here we demonstrate the model architecture ablation by showing the corresponding boxplots in Fig. \ref{fig:ex5}. {The increasing number of steps leads to proportionally higher VRAM consumption and inference time. Specifically, VRAM usage ranges from approximately 8GB (4 steps) to 24GB (12 steps), with corresponding inference times varying from 0.40s to 1.12s.} Apart from the main experiments shown previously, this experiment can be regarded as a minor ablation study, aiming to strike a balance between computational precision and inference speed.

\section{Discussion}
In this study, we introduced RIIR, a deep learning-based medical image registration method that leverages recurrent inferences as a meta-learning strategy. We extended Recurrent Inference Machines (RIMs) to the image registration problem, which has no explicit forward models. {Given the absence of explicit forward models, our approach can be viewed as a case of amortized iterative inference, where the network learns to progressively refine the registration.} RIIR was extensively evaluated on public brain MR, {lung CT}, and in-house quantitative cardiac MR datasets, and demonstrated consistently improved performance over established deep learning models, both in one-step and iterative settings. Additionally, our ablation study confirmed the importance of incorporating hidden states within the RIM-based framework. 

The acclaimed improvement in registration performance is especially pronounced in scenarios with limited training data, as demonstrated in Figs. \ref{fig:ex1_oasis}, \ref{fig:ex1_nlst}, and \ref{fig:ex1_msasha}. In particular, RIIR achieved superior average evaluation metrics with lower variance. {This performance advantage was particularly pronounced in the NLST experiments, where large deformations need to be estimated with very limited training data. Most baseline models required substantially more training data to achieve comparable performance levels, with the exception of L2O, which benefits from additional input modalities such as dynamically sampled coordinates and MIND features.} Though both GraDIRN and RIIR are iterative methods that use similarity gradients, GraDIRN isolates the update of explicit $\mathcal{L}_{\text{sim}}$ and deep learning-based $\mathcal{L}_{\text{reg}}$ with no internal states, potentially resulting in worse generalization and slower convergence compared to RIIR. In particular, GraDIRN initialized the deformation field randomly by default, which could lead to optimization difficulties when the data were extremely limited during training. RCVM achieves lower folding rates ($|J_{\phi}|_{\leq 0}$) through its composed deformation strategy, where the final transformation is obtained by composing multiple smaller deformations. Although this approach effectively prevents topology-breaking deformations, it comes at the cost of an increased {number of parameters}. {While the training and inference times on the OASIS dataset are the second longest among learning-based methods (after L2O), RCVM exhibits comparable runtime with RIIR for both 3D datasets, and shows faster inference time on the 2D dataset (mSASHA).}  {Notably, LapIRN, as a single-pass multi-resolution method, shows remarkable advantages in both VRAM efficiency and inference speed}.

Although hidden states were used in the original RIM and later work \citep{putzky2017recurrent,putzky2019rim,sabidussi2021recurrent,sabidussi2023dtirim}, their impact on the optimization of RIM-based methods has not been investigated in detail. Our second experiment investigates the impact of these hidden states within RIIR. Our findings reveal that the presence of hidden states, as proposed in the original RIM work~\citep{putzky2017recurrent}, contributes positively to the performance of our model, as shown by the quantitative results in Fig. \ref{fig:ex4_box_dice_nlst}. {Unlike L2O which operates on full resolution features at the output, our implementation of hidden states at downsampled resolution leads to minimal additional memory overhead}.

The ablation study in input combinations (Experiment 3) demonstrates that, in scenarios with limited data, {RIIR with images gradient input achieves superior registration performance in anatomical evaluation metrics, as shown in Fig. {\ref{fig:ex3_combined}}. This improvement is particularly evident in the NLST dataset, where RIIR achieves lower TRE and smoother deformation fields. Although including gradient input can be considered to offer additional information, its impact on regularization varies between datasets (Table {\ref{tab:5percent}}). This could possibly be due to the different similarity losses and registration objectives (intra-subject for NLST versus inter-subject for OASIS)}.

{The regularization analysis reveals that RIIR's performance remains dependent on both the choice and weight of regularization terms. As shown in Fig. {\ref{fig:reg_oasis}} and Fig. {\ref{fig:reg_nlst}}, different regularization strategies lead to varying trade-offs between registration accuracy and deformation regularity. This suggests potential future improvements by incorporating adaptive regularization schemes} \citep{hoopes2021hypermorph,mok2021conditional,reithmeir2024learning}. {Such extensions could enhance RIIR's robustness across different clinical scenarios without manual parameter tuning.}


The superior registration performance and data efficiency of RIIR suggest its potential for applications in medical image registration. However, it is necessary to acknowledge the current limitations, to further enhance the framework in future work. {From an architectural perspective, RIIR employs a relatively conventional design that lacks multi-resolution capability, which has proven effective in methods like LapIRN. The simple convolutional GRU structure could also be enhanced with modern components such as dilated convolutions for larger receptive fields or attention mechanisms for better feature extraction. Another significant limitation lies in GPU memory consumption. Despite having fewer parameters than LapIRN and RCVM, RIIR requires more VRAM due to its recurrent nature, and this memory usage increases linearly with the number of inference steps, as demonstrated in Experiment 5. To better adapt to downstream tasks, potential improvements could include semi-supervised strategy, instance optimization and adaptive regularization strategies, which could enhance the flexibility of the model in different clinical scenarios}.

\section{{Conclusion}}
In conclusion, we present RIIR, a novel recurrent deep-learning framework for medical image registration. RIIR significantly extends the concept of recurrent inference machines for inverse problem solving, to high-dimensional optimization challenges with no closed-form forward models. Meanwhile, RIIR distinguishes itself from previous iterative methods by integrating implicit regularization with explicit loss gradients. Our experiments across diverse medical image datasets demonstrated RIIR's superior accuracy and data efficiency. We also empirically demonstrated the effectiveness of its architectural design and the value of hidden states, significantly enhancing both registration accuracy and data efficiency. RIIR is shown to be an effective and generalizable tool for medical image registration, and potentially extends to other high-dimensional optimization problems.



\section*{Acknowledgments}
This work was partly supported by the TU Delft AI Initiative, Amazon Research Awards, the Dutch Research Council (NWO), and the National Heart, Lung, and Blood Institute, National Institutes of Health by the Division of Intramural Research (Z1A-HL006214).
\bibliography{latex/sample}

\begin{thebibliography}{78}
\providecommand{\natexlab}[1]{#1}
\providecommand{\url}[1]{\texttt{#1}}
\expandafter\ifx\csname urlstyle\endcsname\relax
  \providecommand{\doi}[1]{doi: #1}\else
  \providecommand{\doi}{doi: \begingroup \urlstyle{rm}\Url}\fi

\bibitem[Aberle et~al.(2011)Aberle, Adams, Berg, Black, Clapp, Fagerstrom, Gareen, Gatsonis, Marcus, Sicks, et~al.]{aberle2011reduced}
Denise~R Aberle, Amanda~M Adams, Christine~D Berg, William~C Black, Jonathan~D Clapp, Richard~M Fagerstrom, Ilana~F Gareen, Constantine Gatsonis, Pamela~M Marcus, JD~Sicks, et~al.
\newblock Reduced lung-cancer mortality with low-dose computed tomographic screening.
\newblock \emph{The New England journal of medicine}, 365\penalty0 (5):\penalty0 395--409, 2011.

\bibitem[Andrychowicz et~al.(2016)Andrychowicz, Denil, Gomez, Hoffman, Pfau, Schaul, Shillingford, and De~Freitas]{andrychowicz2016learning}
Marcin Andrychowicz, Misha Denil, Sergio Gomez, Matthew~W Hoffman, David Pfau, Tom Schaul, Brendan Shillingford, and Nando De~Freitas.
\newblock Learning to learn by gradient descent by gradient descent.
\newblock \emph{Advances in Neural Information Processing Systems}, 29, 2016.

\bibitem[Avants et~al.(2008)Avants, Epstein, Grossman, and Gee]{avants2008symmetric}
Brian~B Avants, Charles~L Epstein, Murray Grossman, and James~C Gee.
\newblock Symmetric diffeomorphic image registration with cross-correlation: evaluating automated labeling of elderly and neurodegenerative brain.
\newblock \emph{Medical Image Analysis}, 12\penalty0 (1):\penalty0 26--41, 2008.

\bibitem[Avants et~al.(2011)Avants, Tustison, Song, Cook, Klein, and Gee]{avants2011reproducible}
Brian~B Avants, Nicholas~J Tustison, Gang Song, Philip~A Cook, Arno Klein, and James~C Gee.
\newblock A reproducible evaluation of ants similarity metric performance in brain image registration.
\newblock \emph{NeuroImage}, 54\penalty0 (3):\penalty0 2033--2044, 2011.

\bibitem[Balakrishnan et~al.(2019)Balakrishnan, Zhao, Sabuncu, Guttag, and Dalca]{balakrishnan2019voxelmorph}
Guha Balakrishnan, Amy Zhao, Mert~R Sabuncu, John Guttag, and Adrian~V Dalca.
\newblock Voxelmorph: a learning framework for deformable medical image registration.
\newblock \emph{IEEE Transactions on Medical Imaging}, 38\penalty0 (8):\penalty0 1788--1800, 2019.

\bibitem[Byrne et~al.(2022)Byrne, Archibald-Heeren, Hu, Teh, Beserminji, Cai, Liu, Yates, Rijken, Collett, and Aland]{byrne2022varian}
Mikel Byrne, Ben Archibald-Heeren, Yunfei Hu, Amy Teh, Rhea Beserminji, Emma Cai, Guilin Liu, Angela Yates, James Rijken, Nick Collett, and Trent Aland.
\newblock Varian ethos online adaptive radiotherapy for prostate cancer: Early results of contouring accuracy, treatment plan quality, and treatment time.
\newblock \emph{Journal of Applied Clinical Medical Physics}, 23\penalty0 (1):\penalty0 e13479, 2022.
\newblock \doi{10.1002/acm2.13479}.

\bibitem[Cao et~al.(2017)Cao, Yang, Zhang, Nie, Kim, Wang, and Shen]{cao2017deformable}
Xiaohuan Cao, Jianhua Yang, Jun Zhang, Dong Nie, Minjeong Kim, Qian Wang, and Dinggang Shen.
\newblock Deformable image registration based on similarity-steered cnn regression.
\newblock In \emph{Medical Image Computing and Computer Assisted Intervention- MICCAI 2017: 20th International Conference, Quebec City, QC, Canada, September 11-13, 2017, Proceedings, Part I 20}, pages 300--308. Springer, 2017.

\bibitem[Chen et~al.(2017)Chen, Hoffman, Colmenarejo, Denil, Lillicrap, Botvinick, and Freitas]{chen2017learning}
Yutian Chen, Matthew~W Hoffman, Sergio~G{\'o}mez Colmenarejo, Misha Denil, Timothy~P Lillicrap, Matt Botvinick, and Nando Freitas.
\newblock Learning to learn without gradient descent by gradient descent.
\newblock In \emph{International Conference on Machine Learning}, pages 748--756. PMLR, 2017.

\bibitem[Chow et~al.(2022)Chow, Hayes, Flewitt, Feuchter, Lydell, Howarth, Pagano, Thompson, Kellman, and White]{chow2022improved}
Kelvin Chow, Genevieve Hayes, Jacqueline~A Flewitt, Patricia Feuchter, Carmen Lydell, Andrew Howarth, Joseph~J Pagano, Richard~B Thompson, Peter Kellman, and James~A White.
\newblock Improved accuracy and precision with three-parameter simultaneous myocardial t1 and t2 mapping using multiparametric sasha.
\newblock \emph{Magnetic Resonance in Medicine}, 87\penalty0 (6):\penalty0 2775--2791, 2022.

\bibitem[Chung et~al.(2014)Chung, Gulcehre, Cho, and Bengio]{chung2014empirical}
Junyoung Chung, Caglar Gulcehre, KyungHyun Cho, and Yoshua Bengio.
\newblock Empirical evaluation of gated recurrent neural networks on sequence modeling.
\newblock \emph{arXiv preprint arXiv:1412.3555}, 2014.

\bibitem[Cotter and Conwell(1990)]{cotter1990fixed}
Neil~E Cotter and Peter~R Conwell.
\newblock Fixed-weight networks can learn.
\newblock In \emph{International Joint Conference on Neural Networks}, pages 553--559. IEEE, 1990.

\bibitem[Dalca et~al.(2019)Dalca, Balakrishnan, Guttag, and Sabuncu]{dalca2019unsupervised}
Adrian~V Dalca, Guha Balakrishnan, John Guttag, and Mert~R Sabuncu.
\newblock Unsupervised learning of probabilistic diffeomorphic registration for images and surfaces.
\newblock \emph{Medical Image Analysis}, 57:\penalty0 226--236, 2019.

\bibitem[De~Vos et~al.(2019)De~Vos, Berendsen, Viergever, Sokooti, Staring, and I{\v{s}}gum]{de2019deep}
Bob~D De~Vos, Floris~F Berendsen, Max~A Viergever, Hessam Sokooti, Marius Staring, and Ivana I{\v{s}}gum.
\newblock A deep learning framework for unsupervised affine and deformable image registration.
\newblock \emph{Medical Image Analysis}, 52:\penalty0 128--143, 2019.

\bibitem[de~Vos et~al.(2020)de~Vos, van~der Velden, Sander, Gilhuijs, Staring, and I{\v{s}}gum]{de2020mutual}
Bob~D de~Vos, Bas~HM van~der Velden, J{\"o}rg Sander, Kenneth~GA Gilhuijs, Marius Staring, and Ivana I{\v{s}}gum.
\newblock Mutual information for unsupervised deep learning image registration.
\newblock In \emph{Medical Imaging 2020: Image Processing}, volume 11313, pages 155--161. SPIE, 2020.

\bibitem[Falta et~al.(2022)Falta, Hansen, and Heinrich]{falta2022learning}
Fenja Falta, Lasse Hansen, and Mattias~P Heinrich.
\newblock Learning iterative optimisation for deformable image registration of lung ct with recurrent convolutional networks.
\newblock In \emph{International Conference on Medical Image Computing and Computer-Assisted Intervention}, pages 301--309. Springer, 2022.

\bibitem[Fechter and Baltas(2020)]{fechter2020one}
Tobias Fechter and Dimos Baltas.
\newblock One-shot learning for deformable medical image registration and periodic motion tracking.
\newblock \emph{IEEE Transactions on Medical Imaging}, 39\penalty0 (7):\penalty0 2506--2517, 2020.

\bibitem[Finn et~al.(2017)Finn, Abbeel, and Levine]{finn2017model}
Chelsea Finn, Pieter Abbeel, and Sergey Levine.
\newblock Model-agnostic meta-learning for fast adaptation of deep networks.
\newblock In \emph{International Conference on Machine Learning}, pages 1126--1135. PMLR, 2017.

\bibitem[Fischer and Modersitzki(2002)]{fischer2002fast}
Bernd Fischer and Jan Modersitzki.
\newblock Fast diffusion registration.
\newblock \emph{Contemporary Mathematics}, 313:\penalty0 117--128, 2002.

\bibitem[Fischer and Modersitzki(2004)]{fischer2004unified}
Bernd Fischer and Jan Modersitzki.
\newblock A unified approach to fast image registration and a new curvature based registration technique.
\newblock \emph{Linear Algebra and its applications}, 380:\penalty0 107--124, 2004.

\bibitem[Haskins et~al.(2020)Haskins, Kruger, and Yan]{haskins2020deep}
Grant Haskins, Uwe Kruger, and Pingkun Yan.
\newblock Deep learning in medical image registration: a survey.
\newblock \emph{Machine Vision and Applications}, 31:\penalty0 1--18, 2020.

\bibitem[Heinrich et~al.(2012)Heinrich, Jenkinson, Bhushan, Matin, Gleeson, Brady, and Schnabel]{heinrich2012mind}
Mattias~P Heinrich, Mark Jenkinson, Manav Bhushan, Tahreema Matin, Fergus~V Gleeson, Michael Brady, and Julia~A Schnabel.
\newblock Mind: Modality independent neighbourhood descriptor for multi-modal deformable registration.
\newblock \emph{Medical image analysis}, 16\penalty0 (7):\penalty0 1423--1435, 2012.

\bibitem[Heinrich et~al.(2015)Heinrich, Handels, and Simpson]{heinrich2015estimating}
Mattias~P Heinrich, Heinz Handels, and Ivor~JA Simpson.
\newblock Estimating large lung motion in copd patients by symmetric regularised correspondence fields.
\newblock In \emph{Medical Image Computing and Computer-Assisted Intervention--MICCAI 2015: 18th International Conference, Munich, Germany, October 5-9, 2015, Proceedings, Part II 18}, pages 338--345. Springer, 2015.

\bibitem[Hering et~al.(2019)Hering, van Ginneken, and Heldmann]{hering2019mlvirnet}
Alessa Hering, Bram van Ginneken, and Stefan Heldmann.
\newblock mlvirnet: Multilevel variational image registration network.
\newblock In \emph{Medical Image Computing and Computer Assisted Intervention--MICCAI 2019: 22nd International Conference, Shenzhen, China, October 13--17, 2019, Proceedings, Part VI 22}, pages 257--265. Springer, 2019.

\bibitem[Hering et~al.(2022)Hering, Hansen, Mok, Chung, Siebert, H{\"a}ger, Lange, Kuckertz, Heldmann, Shao, et~al.]{hering2022learn2reg}
Alessa Hering, Lasse Hansen, Tony~CW Mok, Albert~CS Chung, Hanna Siebert, Stephanie H{\"a}ger, Annkristin Lange, Sven Kuckertz, Stefan Heldmann, Wei Shao, et~al.
\newblock Learn2reg: comprehensive multi-task medical image registration challenge, dataset and evaluation in the era of deep learning.
\newblock \emph{IEEE Transactions on Medical Imaging}, 42\penalty0 (3):\penalty0 697--712, 2022.

\bibitem[Hoopes et~al.(2021)Hoopes, Hoffmann, Fischl, Guttag, and Dalca]{hoopes2021hypermorph}
Andrew Hoopes, Malte Hoffmann, Bruce Fischl, John Guttag, and Adrian~V Dalca.
\newblock Hypermorph: Amortized hyperparameter learning for image registration.
\newblock In \emph{Information Processing in Medical Imaging: 27th International Conference, IPMI 2021, Virtual Event, June 28--June 30, 2021, Proceedings 27}, pages 3--17. Springer, 2021.

\bibitem[Hospedales et~al.(2021)Hospedales, Antoniou, Micaelli, and Storkey]{hospedales2021meta}
Timothy Hospedales, Antreas Antoniou, Paul Micaelli, and Amos Storkey.
\newblock Meta-learning in neural networks: A survey.
\newblock \emph{IEEE Transactions on Pattern Analysis and Machine Intelligence}, 44\penalty0 (9):\penalty0 5149--5169, 2021.

\bibitem[Huizinga et~al.(2016)Huizinga, Poot, Guyader, Klaassen, Coolen, {van Kranenburg}, {van Geuns}, Uitterdijk, Polfliet, Vandemeulebroucke, Leemans, Niessen, and Klein]{huizinga2016pca}
W.~Huizinga, D.H.J. Poot, J.-M. Guyader, R.~Klaassen, B.F. Coolen, M.~{van Kranenburg}, R.J.M. {van Geuns}, A.~Uitterdijk, M.~Polfliet, J.~Vandemeulebroucke, A.~Leemans, W.J. Niessen, and S.~Klein.
\newblock Pca-based groupwise image registration for quantitative mri.
\newblock \emph{Medical Image Analysis}, 29:\penalty0 65--78, 2016.
\newblock ISSN 1361-8415.
\newblock \doi{https://doi.org/10.1016/j.media.2015.12.004}.
\newblock URL \url{https://www.sciencedirect.com/science/article/pii/S1361841515001851}.

\bibitem[Isensee et~al.(2021)Isensee, Jaeger, Kohl, Petersen, and Maier-Hein]{isensee2021nnu}
Fabian Isensee, Paul~F Jaeger, Simon~AA Kohl, Jens Petersen, and Klaus~H Maier-Hein.
\newblock nnu-net: a self-configuring method for deep learning-based biomedical image segmentation.
\newblock \emph{Nature methods}, 18\penalty0 (2):\penalty0 203--211, 2021.

\bibitem[Jin et~al.(2021)Jin, Yu, Ke, Ding, Yi, Jiang, Duan, Tang, Chang, Wu, Gao, and Li]{jin2021predicting}
Cheng Jin, Heng Yu, Jia Ke, Peirong Ding, Yongju Yi, Xiaofeng Jiang, Xin Duan, Jinghua Tang, Daniel~T Chang, Xiaojian Wu, Feng Gao, and Ruijiang Li.
\newblock Predicting treatment response from longitudinal images using multi-task deep learning.
\newblock \emph{Nature Communications}, 12\penalty0 (1):\penalty0 1851, 2021.

\bibitem[Kanter and Lellmann(2022)]{kanter2022flexible}
Frederic Kanter and Jan Lellmann.
\newblock A flexible meta learning model for image registration.
\newblock In Ender Konukoglu, Bjoern Menze, Archana Venkataraman, Christian Baumgartner, Qi~Dou, and Shadi Albarqouni, editors, \emph{Proceedings of The 5th Medical Imaging with Deep Learning}, volume 172 of \emph{Proceedings of Machine Learning Research}, pages 638--652. PMLR, 06--08 Jul 2022.
\newblock URL \url{https://proceedings.mlr.press/v172/kanter22a.html}.

\bibitem[Karkalousos et~al.(2022)Karkalousos, Noteboom, Hulst, Vos, and Caan]{karkalousos2022assessment}
Dimitrios Karkalousos, Samantha Noteboom, Hanneke~E Hulst, Franciscus~M Vos, and Matthan~WA Caan.
\newblock Assessment of data consistency through cascades of independently recurrent inference machines for fast and robust accelerated mri reconstruction.
\newblock \emph{Physics in Medicine \& Biology}, 67\penalty0 (12):\penalty0 124001, 2022.

\bibitem[King et~al.(2010)King, Rhode, Ma, Yao, Jansen, Razavi, and Penney]{king2010registering}
Andrew~P King, Kawal~S Rhode, Y~Ma, Cheng Yao, Christian Jansen, Reza Razavi, and Graeme~P Penney.
\newblock Registering preprocedure volumetric images with intraprocedure 3-d ultrasound using an ultrasound imaging model.
\newblock \emph{IEEE Transactions on Medical Imaging}, 29\penalty0 (3):\penalty0 924--937, 2010.

\bibitem[Kingma and Ba(2015)]{adampaper}
Diederik~P. Kingma and Jimmy Ba.
\newblock Adam: {A} method for stochastic optimization.
\newblock In Yoshua Bengio and Yann LeCun, editors, \emph{3rd International Conference on Learning Representations, {ICLR} 2015, San Diego, CA, USA, May 7-9, 2015, Conference Track Proceedings}, 2015.
\newblock URL \url{http://arxiv.org/abs/1412.6980}.

\bibitem[Klein et~al.(2007)Klein, Staring, and Pluim]{klein2007evaluation}
Stefan Klein, Marius Staring, and Josien~PW Pluim.
\newblock Evaluation of optimization methods for nonrigid medical image registration using mutual information and b-splines.
\newblock \emph{IEEE Transactions on Image Processing}, 16\penalty0 (12):\penalty0 2879--2890, 2007.

\bibitem[Klein et~al.(2009)Klein, Staring, Murphy, Viergever, and Pluim]{klein2009elastix}
Stefan Klein, Marius Staring, Keelin Murphy, Max~A Viergever, and Josien~PW Pluim.
\newblock Elastix: a toolbox for intensity-based medical image registration.
\newblock \emph{IEEE transactions on medical imaging}, 29\penalty0 (1):\penalty0 196--205, 2009.

\bibitem[Kumaresan and Radhakrishnan(1996)]{kumaresan1996importance}
S~Kumaresan and S~Radhakrishnan.
\newblock Importance of partitioning membranes of the brain and the influence of the neck in head injury modelling.
\newblock \emph{Medical and Biological Engineering and Computing}, 34:\penalty0 27--32, 1996.

\bibitem[Lai-Fook and Hyatt(2000)]{lai2000effects}
Stephen~J Lai-Fook and Robert~E Hyatt.
\newblock Effects of age on elastic moduli of human lungs.
\newblock \emph{Journal of applied physiology}, 89\penalty0 (1):\penalty0 163--168, 2000.

\bibitem[Liu et~al.(2021)Liu, Li, Fan, Zhao, Huang, and Luo]{liu2021learning}
Risheng Liu, Zi~Li, Xin Fan, Chenying Zhao, Hao Huang, and Zhongxuan Luo.
\newblock Learning deformable image registration from optimization: perspective, modules, bilevel training and beyond.
\newblock \emph{IEEE Transactions on Pattern Analysis and Machine Intelligence}, 44\penalty0 (11):\penalty0 7688--7704, 2021.

\bibitem[L{\o}nning et~al.(2019)L{\o}nning, Putzky, Sonke, Reneman, Caan, and Welling]{lonning2019recurrent}
Kai L{\o}nning, Patrick Putzky, Jan-Jakob Sonke, Liesbeth Reneman, Matthan~WA Caan, and Max Welling.
\newblock Recurrent inference machines for reconstructing heterogeneous mri data.
\newblock \emph{Medical Image Analysis}, 53:\penalty0 64--78, 2019.

\bibitem[Marcus et~al.(2007)Marcus, Wang, Parker, Csernansky, Morris, and Buckner]{marcus2007open}
Daniel~S Marcus, Tracy~H Wang, Jamie Parker, John~G Csernansky, John~C Morris, and Randy~L Buckner.
\newblock Open access series of imaging studies (oasis): cross-sectional mri data in young, middle aged, nondemented, and demented older adults.
\newblock \emph{Journal of Cognitive Neuroscience}, 19\penalty0 (9):\penalty0 1498--1507, 2007.

\bibitem[Marino et~al.(2018)Marino, Yue, and Mandt]{marino2018iterative}
Joe Marino, Yisong Yue, and Stephan Mandt.
\newblock Iterative amortized inference.
\newblock In \emph{International Conference on Machine Learning}, pages 3403--3412. PMLR, 2018.

\bibitem[Messroghli et~al.(2004)Messroghli, Radjenovic, Kozerke, Higgins, Sivananthan, and Ridgway]{messroghli2004modified}
Daniel~R Messroghli, Aleksandra Radjenovic, Sebastian Kozerke, David~M Higgins, Mohan~U Sivananthan, and John~P Ridgway.
\newblock Modified look-locker inversion recovery (molli) for high-resolution t1 mapping of the heart.
\newblock \emph{Magnetic Resonance in Medicine}, 52\penalty0 (1):\penalty0 141--146, 2004.

\bibitem[Miao et~al.(2016)Miao, Wang, and Liao]{miao2016cnn}
Shun Miao, Z~Jane Wang, and Rui Liao.
\newblock A cnn regression approach for real-time 2d/3d registration.
\newblock \emph{IEEE Transactions on Medical Imaging}, 35\penalty0 (5):\penalty0 1352--1363, 2016.

\bibitem[Modi et~al.(2021)Modi, Lanusse, Seljak, Spergel, and Perreault-Levasseur]{modi2021cosmicrim}
Chirag Modi, Fran{\c{c}}ois Lanusse, Uro{\v{s}} Seljak, David~N Spergel, and Laurence Perreault-Levasseur.
\newblock Cosmicrim: reconstructing early universe by combining differentiable simulations with recurrent inference machines.
\newblock \emph{arXiv preprint arXiv:2104.12864}, 2021.

\bibitem[Mok and Chung(2020)]{mok2020large}
Tony~CW Mok and Albert~CS Chung.
\newblock Large deformation diffeomorphic image registration with laplacian pyramid networks.
\newblock In \emph{Medical Image Computing and Computer Assisted Intervention--MICCAI 2020: 23rd International Conference, Lima, Peru, October 4--8, 2020, Proceedings, Part III 23}, pages 211--221. Springer, 2020.

\bibitem[Mok and Chung(2021)]{mok2021conditional}
Tony~CW Mok and Albert~CS Chung.
\newblock Conditional deformable image registration with convolutional neural network.
\newblock In \emph{Medical Image Computing and Computer Assisted Intervention--MICCAI 2021: 24th International Conference, Strasbourg, France, September 27--October 1, 2021, Proceedings, Part IV 24}, pages 35--45. Springer, 2021.

\bibitem[Morningstar et~al.(2019)Morningstar, Levasseur, Hezaveh, Blandford, Marshall, Putzky, Rueter, Wechsler, and Welling]{morningstar2019data}
Warren~R Morningstar, Laurence~Perreault Levasseur, Yashar~D Hezaveh, Roger Blandford, Phil Marshall, Patrick Putzky, Thomas~D Rueter, Risa Wechsler, and Max Welling.
\newblock Data-driven reconstruction of gravitationally lensed galaxies using recurrent inference machines.
\newblock \emph{The Astrophysical Journal}, 883\penalty0 (1):\penalty0 14, 2019.
\newblock \doi{10.1109/TIP.2007.909412}.

\bibitem[Muckley et~al.(2021)Muckley, Riemenschneider, Radmanesh, Kim, Jeong, Ko, Jun, Shin, Hwang, Mostapha, Arberet, Nickel, Ramzi, Ciuciu, Starck, Teuwen, Karkalousos, Zhang, Sriram, Huang, Yakubova, Lui, and Knoll]{muckley2021results}
Matthew~J. Muckley, Bruno Riemenschneider, Alireza Radmanesh, Sunwoo Kim, Geunu Jeong, Jingyu Ko, Yohan Jun, Hyungseob Shin, Dosik Hwang, Mahmoud Mostapha, Simon Arberet, Dominik Nickel, Zaccharie Ramzi, Philippe Ciuciu, Jean-Luc Starck, Jonas Teuwen, Dimitrios Karkalousos, Chaoping Zhang, Anuroop Sriram, Zhengnan Huang, Nafissa Yakubova, Yvonne~W. Lui, and Florian Knoll.
\newblock Results of the 2020 fastmri challenge for machine learning mr image reconstruction.
\newblock \emph{IEEE Transactions on Medical Imaging}, 40\penalty0 (9):\penalty0 2306--2317, 2021.
\newblock \doi{10.1109/TMI.2021.3075856}.

\bibitem[Ntatsis et~al.(2023)Ntatsis, Dekker, Van Der~Valk, Birdsong, Zukiczukic, Klein, Staring, and Mccormick]{ntatsis2023itk}
Konstantinos Ntatsis, Niels Dekker, Viktor Van Der~Valk, Tom Birdsong, D~Zukiczukic, Stefan Klein, Marius Staring, and Matthew Mccormick.
\newblock itk-elastix: Medical image registration in python.
\newblock In \emph{Proceedings of the 22nd Python in Science Conference}, pages 101--105, 2023.

\bibitem[Oliveira and Tavares(2014)]{oliveira2014medical}
Francisco~PM Oliveira and Joao Manuel~RS Tavares.
\newblock Medical image registration: a review.
\newblock \emph{Computer Methods in Biomechanics and Biomedical Engineering}, 17\penalty0 (2):\penalty0 73--93, 2014.

\bibitem[Paszke et~al.(2017)Paszke, Gross, Chintala, Chanan, Yang, DeVito, Lin, Desmaison, Antiga, and Lerer]{paszke2017automatic}
Adam Paszke, Sam Gross, Soumith Chintala, Gregory Chanan, Edward Yang, Zachary DeVito, Zeming Lin, Alban Desmaison, Luca Antiga, and Adam Lerer.
\newblock Automatic differentiation in pytorch.
\newblock 2017.

\bibitem[Putzky and Welling(2017)]{putzky2017recurrent}
Patrick Putzky and Max Welling.
\newblock Recurrent inference machines for solving inverse problems.
\newblock \emph{arXiv preprint arXiv:1706.04008}, 2017.

\bibitem[Putzky et~al.(2019)Putzky, Karkalousos, Teuwen, Miriakov, Bakker, Caan, and Welling]{putzky2019rim}
Patrick Putzky, Dimitrios Karkalousos, Jonas Teuwen, Nikita Miriakov, Bart Bakker, Matthan Caan, and Max Welling.
\newblock i-rim applied to the fastmri challenge.
\newblock \emph{arXiv preprint arXiv:1910.08952}, 2019.

\bibitem[Qiu et~al.(2021)Qiu, Qin, Schuh, Hammernik, and Rueckert]{qiu2021learning}
Huaqi Qiu, Chen Qin, Andreas Schuh, Kerstin Hammernik, and Daniel Rueckert.
\newblock Learning diffeomorphic and modality-invariant registration using b-splines.
\newblock In \emph{Medical Imaging with Deep Learning}, 2021.

\bibitem[Qiu et~al.(2022)Qiu, Hammernik, Qin, Chen, and Rueckert]{qiu2022embedding}
Huaqi Qiu, Kerstin Hammernik, Chen Qin, Chen Chen, and Daniel Rueckert.
\newblock Embedding gradient-based optimization in image registration networks.
\newblock In \emph{Medical Image Computing and Computer-Assisted Intervention--MICCAI 2022}, pages 56--65. Springer, 2022.

\bibitem[Reithmeir et~al.(2024)Reithmeir, Schnabel, and Zimmer]{reithmeir2024learning}
Anna Reithmeir, Julia~A Schnabel, and Veronika~A Zimmer.
\newblock Learning physics-inspired regularization for medical image registration with hypernetworks.
\newblock In \emph{Medical Imaging 2024: Image Processing}, volume 12926, pages 625--635. SPIE, 2024.

\bibitem[Roh{\'e} et~al.(2017)Roh{\'e}, Datar, Heimann, Sermesant, and Pennec]{rohe2017svf}
Marc-Michel Roh{\'e}, Manasi Datar, Tobias Heimann, Maxime Sermesant, and Xavier Pennec.
\newblock Svf-net: learning deformable image registration using shape matching.
\newblock In \emph{Medical Image Computing and Computer Assisted Intervention- MICCAI 2017: 20th International Conference, Quebec City, QC, Canada, September 11-13, 2017, Proceedings, Part I 20}, pages 266--274. Springer, 2017.

\bibitem[Ronneberger et~al.(2015)Ronneberger, Fischer, and Brox]{ronneberger2015unet}
Olaf Ronneberger, Philipp Fischer, and Thomas Brox.
\newblock U-net: Convolutional networks for biomedical image segmentation.
\newblock In \emph{Medical Image Computing and Computer-Assisted Intervention--MICCAI 2015: 18th International Conference, Munich, Germany, October 5-9, 2015, Proceedings, Part III 18}, pages 234--241. Springer, 2015.

\bibitem[Rueckert and Schnabel(2019)]{rueckert2019model}
Daniel Rueckert and Julia~A Schnabel.
\newblock Model-based and data-driven strategies in medical image computing.
\newblock \emph{Proceedings of the IEEE}, 108\penalty0 (1):\penalty0 110--124, 2019.

\bibitem[Sabidussi et~al.(2021)Sabidussi, Klein, Caan, Bazrafkan, den Dekker, Sijbers, Niessen, and Poot]{sabidussi2021recurrent}
Emanoel~R Sabidussi, Stefan Klein, Matthan~WA Caan, Shabab Bazrafkan, Arnold~J den Dekker, Jan Sijbers, Wiro~J Niessen, and Dirk~HJ Poot.
\newblock Recurrent inference machines as inverse problem solvers for mr relaxometry.
\newblock \emph{Medical Image Analysis}, 74:\penalty0 102220, 2021.

\bibitem[Sabidussi et~al.(2023)Sabidussi, Klein, Jeurissen, and Poot]{sabidussi2023dtirim}
ER~Sabidussi, S~Klein, B~Jeurissen, and DHJ Poot.
\newblock dti{RIM}: A generalisable deep learning method for diffusion tensor imaging.
\newblock \emph{NeuroImage}, 269:\penalty0 119900, 2023.

\bibitem[Sandk{\"u}hler et~al.(2019)Sandk{\"u}hler, Andermatt, Bauman, Nyilas, Jud, and Cattin]{sandkuhler2019recurrent}
Robin Sandk{\"u}hler, Simon Andermatt, Grzegorz Bauman, Sylvia Nyilas, Christoph Jud, and Philippe~C Cattin.
\newblock Recurrent registration neural networks for deformable image registration.
\newblock \emph{Advances in Neural Information Processing Systems}, 32, 2019.

\bibitem[Sauer(2006)]{sauer2006image}
Frank Sauer.
\newblock Image registration: enabling technology for image guided surgery and therapy.
\newblock In \emph{2005 IEEE Engineering in Medicine and Biology 27th Annual Conference}, pages 7242--7245. IEEE, 2006.

\bibitem[Schmidhuber(1993)]{schmidhuber1993neural}
J{\"u}rgen Schmidhuber.
\newblock A neural network that embeds its own meta-levels.
\newblock In \emph{IEEE International Conference on Neural Networks}, pages 407--412. IEEE, 1993.

\bibitem[Shi et~al.(2015)Shi, Chen, Wang, Yeung, Wong, and Woo]{shi2015convolutional}
Xingjian Shi, Zhourong Chen, Hao Wang, Dit-Yan Yeung, Wai-Kin Wong, and Wang-chun Woo.
\newblock Convolutional lstm network: A machine learning approach for precipitation nowcasting.
\newblock \emph{Advances in Neural Information Processing Systems}, 28, 2015.

\bibitem[Sotiras et~al.(2013)Sotiras, Davatzikos, and Paragios]{sotiras2013deformable}
Aristeidis Sotiras, Christos Davatzikos, and Nikos Paragios.
\newblock Deformable medical image registration: A survey.
\newblock \emph{IEEE Transactions on Medical Imaging}, 32\penalty0 (7):\penalty0 1153--1190, 2013.

\bibitem[Staring et~al.(2009)Staring, van~der Heide, Klein, Viergever, and Pluim]{staring2009registration}
Marius Staring, Uulke~A van~der Heide, Stefan Klein, Max~A Viergever, and Josien~PW Pluim.
\newblock Registration of cervical mri using multifeature mutual information.
\newblock \emph{IEEE Transactions on Medical Imaging}, 28\penalty0 (9):\penalty0 1412--1421, 2009.

\bibitem[Studholme et~al.(1999)Studholme, Hill, and Hawkes]{studholme1999overlap}
Colin Studholme, Derek~LG Hill, and David~J Hawkes.
\newblock An overlap invariant entropy measure of 3d medical image alignment.
\newblock \emph{Pattern Recognition}, 32\penalty0 (1):\penalty0 71--86, 1999.

\bibitem[Th{\'e}venaz and Unser(2000)]{thevenaz2000optimization}
Philippe Th{\'e}venaz and Michael Unser.
\newblock Optimization of mutual information for multiresolution image registration.
\newblock \emph{IEEE Transactions on Image Processing}, 9\penalty0 (12):\penalty0 2083--2099, 2000.

\bibitem[van Harten et~al.(2023)van Harten, Van~Herten, Stoker, and Isgum]{van2023deformable}
Louis van Harten, Rudolf Leonardus~Mirjam Van~Herten, Jaap Stoker, and Ivana Isgum.
\newblock Deformable image registration with geometry-informed implicit neural representations.
\newblock In \emph{Medical Imaging with Deep Learning}, 2023.

\bibitem[Wolterink et~al.(2022)Wolterink, Zwienenberg, and Brune]{wolterink2022implicit}
Jelmer~M Wolterink, Jesse~C Zwienenberg, and Christoph Brune.
\newblock Implicit neural representations for deformable image registration.
\newblock In \emph{Medical Imaging with Deep Learning}, pages 1349--1359. PMLR, 2022.

\bibitem[Xu et~al.(2021)Xu, Chen, Chen, Chen, and Sun]{xu2021multi}
Junshen Xu, Eric~Z Chen, Xiao Chen, Terrence Chen, and Shanhui Sun.
\newblock Multi-scale neural {ode}s for 3d medical image registration.
\newblock In \emph{Medical Image Computing and Computer Assisted Intervention--MICCAI 2021: 24th International Conference, Strasbourg, France, September 27--October 1, 2021, Proceedings, Part IV 24}, pages 213--223. Springer, 2021.

\bibitem[Yang et~al.(2016)Yang, Kwitt, and Niethammer]{yang2016fast}
Xiao Yang, Roland Kwitt, and Marc Niethammer.
\newblock Fast predictive image registration.
\newblock In \emph{Deep Learning and Data Labeling for Medical Applications: First International Workshop, LABELS 2016, and Second International Workshop, DLMIA 2016, Held in Conjunction with MICCAI 2016, Athens, Greece, October 21, 2016, Proceedings 1}, pages 48--57. Springer, 2016.

\bibitem[Yang et~al.(2017)Yang, Kwitt, Styner, and Niethammer]{YANG2017378}
Xiao Yang, Roland Kwitt, Martin Styner, and Marc Niethammer.
\newblock Quicksilver: Fast predictive image registration – a deep learning approach.
\newblock \emph{NeuroImage}, 158:\penalty0 378--396, 2017.
\newblock ISSN 1053-8119.
\newblock \doi{10.1016/j.NeuroImage.2017.07.008}.

\bibitem[Younger et~al.(1999)Younger, Conwell, and Cotter]{younger1999fixed}
A~Steven Younger, Peter~R Conwell, and Neil~E Cotter.
\newblock Fixed-weight on-line learning.
\newblock \emph{IEEE Transactions on Neural Networks}, 10\penalty0 (2):\penalty0 272--283, 1999.

\bibitem[Zbontar et~al.(2018)Zbontar, Knoll, Sriram, Muckley, Bruno, Defazio, Parente, Geras, Katsnelson, Chandarana, Zhang, Drozdzal, Romero, Rabbat, Vincent, Pinkerton, Wang, Yakubova, Owens, Zitnick, Recht, Sodickson, and Lui]{zbontar2018fastmri}
Jure Zbontar, Florian Knoll, Anuroop Sriram, Matthew~J. Muckley, Mary Bruno, Aaron Defazio, Marc Parente, Krzysztof~J. Geras, Joe Katsnelson, Hersh Chandarana, Zizhao Zhang, Michal Drozdzal, Adriana Romero, Michael~G. Rabbat, Pascal Vincent, James Pinkerton, Duo Wang, Nafissa Yakubova, Erich Owens, C.~Lawrence Zitnick, Michael~P. Recht, Daniel~K. Sodickson, and Yvonne~W. Lui.
\newblock fastmri: An open dataset and benchmarks for accelerated mri.
\newblock \emph{arXiv preprint arXiv:1811.08839}, 2018.

\bibitem[Zhang et~al.(2021)Zhang, Pei, and Zha]{zhang2021learning}
Yungeng Zhang, Yuru Pei, and Hongbin Zha.
\newblock Learning dual transformer network for diffeomorphic registration.
\newblock In \emph{Medical Image Computing and Computer Assisted Intervention--MICCAI 2021: 24th International Conference, Strasbourg, France, September 27--October 1, 2021, Proceedings, Part IV 24}, pages 129--138. Springer, 2021.

\bibitem[Zhao et~al.(2019)Zhao, Dong, Chang, and Xu]{zhao2019recursive}
Shengyu Zhao, Yue Dong, Eric~I Chang, and Yan Xu.
\newblock Recursive cascaded networks for unsupervised medical image registration.
\newblock In \emph{Proceedings of the IEEE/CVF International Conference on Computer Vision}, pages 10600--10610, 2019.

\end{thebibliography}
\section*{Supplementary Materials}
\subsection*{Additional Results of Experiment 1}
\begin{figure}[htb!]
    \centering
    \includegraphics[width=0.65\linewidth]{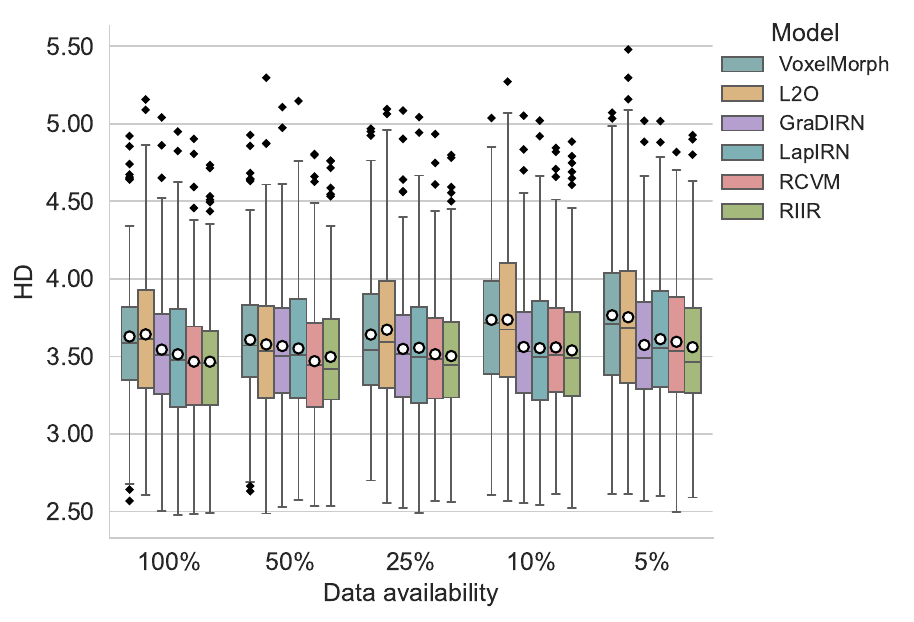}
    \caption{{Supplementary results of Experiment 1 on test OASIS data, showing Hausdorff Distance (HD) comparison between different methods. Lower HD values indicate better registration accuracy and structural preservation.}}
    \label{ex_1_hd_si}
\end{figure}

\begin{figure}[h!]
    \centering
    \includegraphics[width=0.65\linewidth]{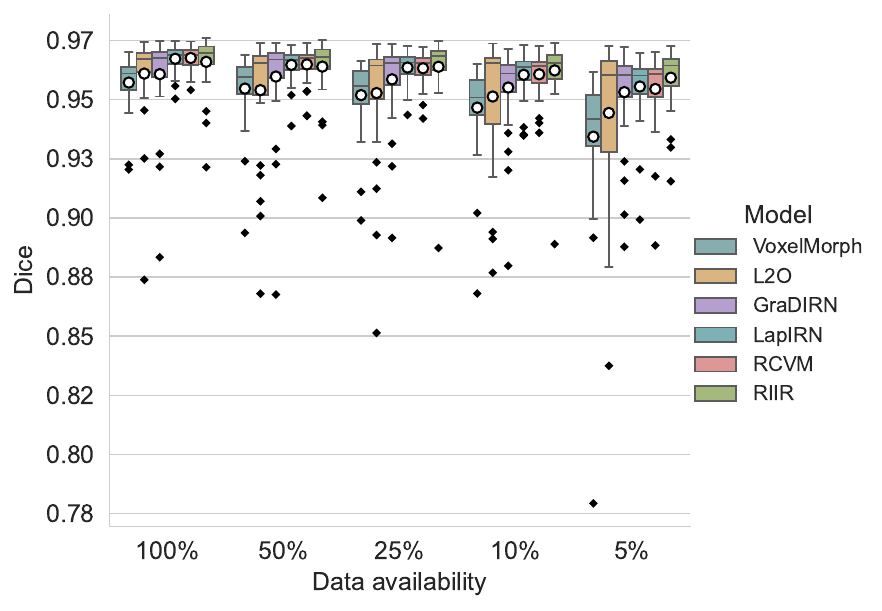}
    \caption{{Supplementary results of Experiment 1 on test NLST data, showing Dice scores for lung lobes between different methods. Higher Dice scores indicate better registration accuracy in aligning lung structures.}}
    \label{ex_1_nlst_dice_si}
\end{figure}

\begin{figure}[h!]
    \centering
    \includegraphics[width=0.65\linewidth]{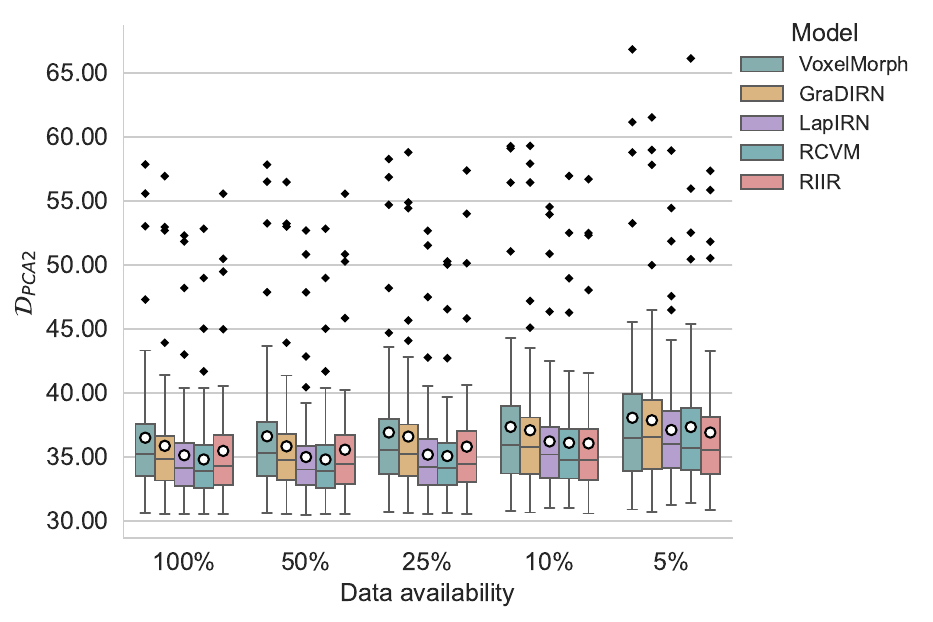}
    \caption{{Supplementary results of Experiment 1 on test mSASHA data, showing $\mathcal{D}_\text{PCA2}$ metric comparison between different methods.}}
    \label{ex_1_pca2_si}
\end{figure}

\begin{figure*}[htb]
    \centering
    \includegraphics[width=0.85\linewidth]{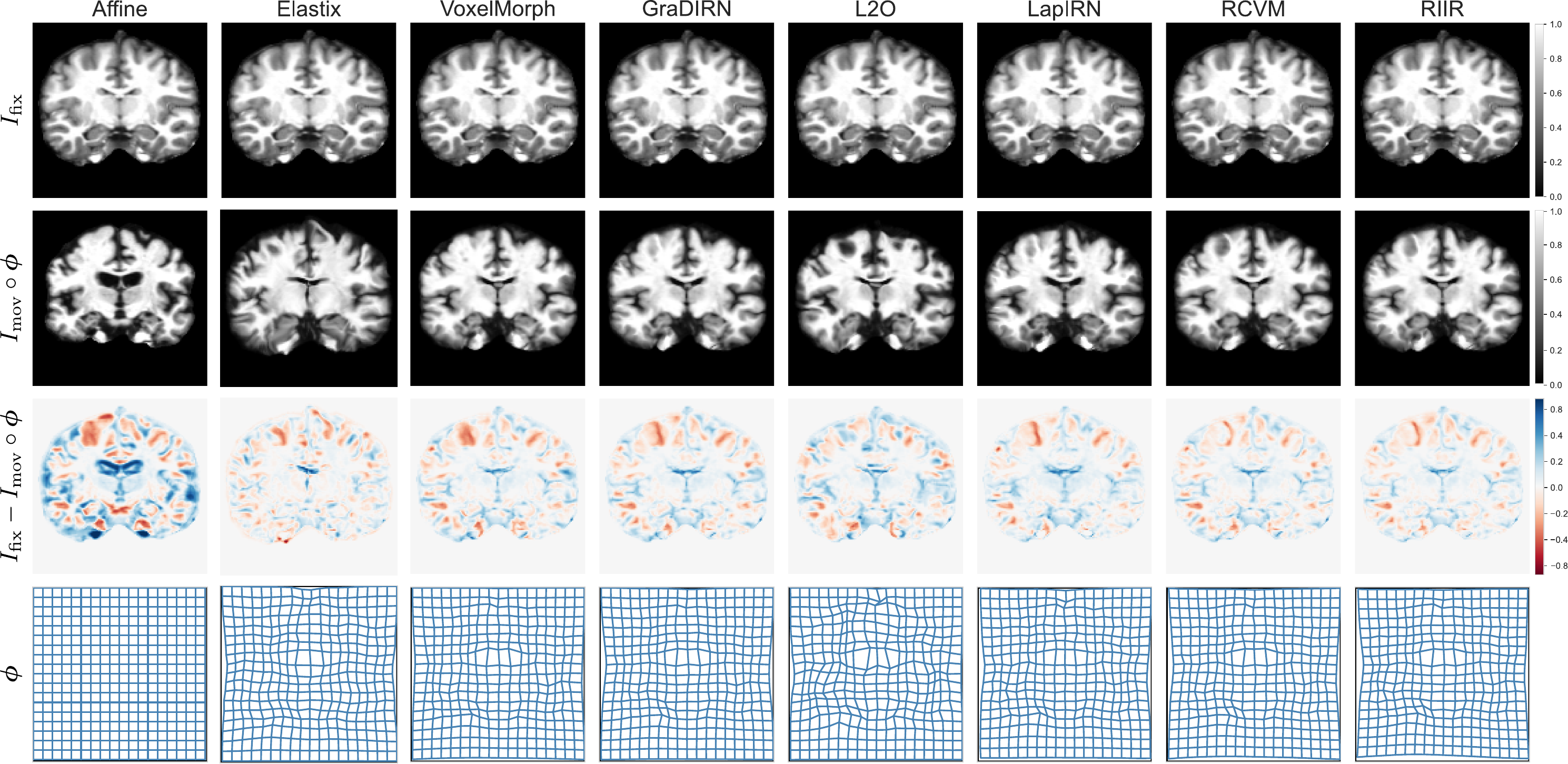}
    \caption{{Qualitative comparison of different registration methods on a representative OASIS test case. The bottom row shows the corresponding deformation fields where `Affine' denotes no further registration applied.}}
    \label{quali}
\end{figure*}

{We show the boxplots of Experiment 1 here in terms of HD on OASIS dataset in Fig. {\ref{ex_1_hd_si}}. Additionally, we present the Dice scores for lung lobes on the NLST dataset in Fig. {\ref{ex_1_nlst_dice_si}}. For mSASHA dataset, we show the $\mathcal{D}_{\text{PCA2}}$ metric in Fig. {\ref{ex_1_pca2_si}}. A qualitative comparison on OASIS is shown in Fig. {\ref{quali}}.}

\subsection*{{Implementation study: Choice of Inner Loss Functions $\mathcal{L}_{\text{inner}}$ for mSASHA}}

\begin{figure}[h!]
    \centering
    \includegraphics[scale=0.5]{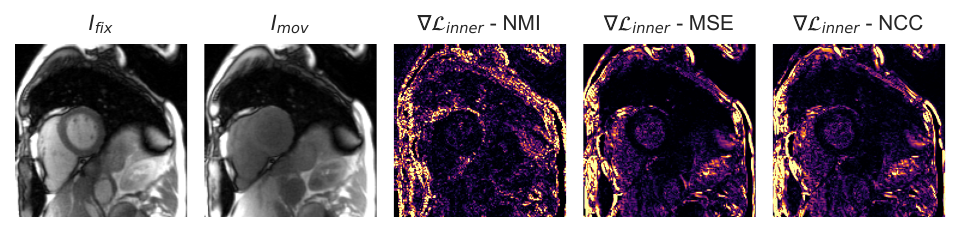}
    \caption{Qualitative check of {gradient information} with an example pair from \textbf{mSASHA}. Three gradient images are taken from the last step of the RIIR inference pipeline at the same epoch.}
    \label{fig:ex_appen_vis}
\end{figure}

\begin{figure}[h!]
    \centering
    \includegraphics[scale = 0.7]{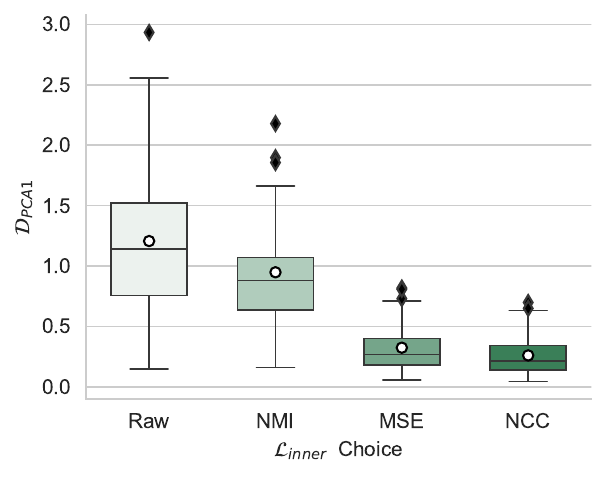}
    \caption{Results evaluated on \textbf{mSASHA} validation set. Raw denotes no registration is implemented to the image series.}
    \label{fig:ex_appen_box}
\end{figure}

Since it has been proposed previously that NMI is a more suitable choice of $\mathcal{L}_{\text{sim}}$ for $\mathcal{L}_{\text{outer}}$ in qMRI registration \citep{de2020mutual} over NCC and MSE, we focused on the choice of $\mathcal{L}_\text{inner}$ to investigate the registration performance and the information propagated through the RIIR {on mSASHA}. In our implementation, \verb|autograd| \citep{paszke2017automatic} were used to calculate $\nabla \mathcal{L}_{\text{inner}}$ for NCC and NMI, while we used the analytical gradient of MSE due to its simplicity. {However, the NMI calculation requires estimating the joint probability distribution using Parzen window (Gaussian kernel) approximation, which involves flattening $I_{\text{fix}}$ and $I_{\text{mov}}\circ \bm{\phi}$ to compute the joint probability.} This approximation results in the scattering pattern shown in Fig. \ref{fig:ex_appen_vis} and influence the performance {when using the RIM input ($[\bm{\phi}_t, \nabla_{\bm{\phi}_t}\mathcal{L}_{\text{inner}}]$)} as shown in Fig. \ref{fig:ex_appen_box}.

\end{document}